\documentclass[12pt]{article}
\usepackage{amsmath,amsfonts}
\usepackage[nosort]{cite}
\newlength{\extraspace}
\newlength{\extraspaces}
 \catcode`\@=11
\def\numberbysection{\@addtoreset{equation}{section}
\def\theequation{\arabic{section}.\arabic{equation}}}
\newcommand{\newsection}[1]{
\vspace{7mm}
\pagebreak[3]
\addtocounter{section}{1}
\setcounter{equation}{0}
\setcounter{subsection}{0}
\setcounter{footnote}{0}
\begin{center}
{\large {\bf \thesection. #1}}
\end{center}
\nopagebreak
\medskip
\nopagebreak
\hspace{3mm}}
\newcommand{\nonu}{\nonumber \\[.5mm]}
\newcommand{\A}{&\!\!\!}

\numberbysection
\begin{document}
\addtolength{\baselineskip}{.7mm}
\thispagestyle{empty}
\begin{flushright}
STUPP--12--209 \\
June, 2012
\end{flushright}
\vspace{20mm}
\begin{center}
{\Large \textbf{Coupling of the BLG theory to \\[2mm]
a conformal supergravity background 
}} \\[20mm]
\textsc{Madoka Nishimura}${}^{\rm a}$
\hspace{1mm} and \hspace{1mm}
\textsc{Yoshiaki Tanii}${}^{\rm b}$\footnote{
\texttt{e-mail: tanii@phy.saitama-u.ac.jp}} \\[7mm]
${}^{\rm a}$\textit{Department of Community Service and Science \\
Tohoku University of Community Service and Science \\
Iimoriyama 3-5-1, Sakata 998-8580, Japan} \\[5mm]
${}^{\rm b}$\textit{Division of Material Science \\ 
Graduate School of Science and Engineering \\
Saitama University, Saitama 338-8570, Japan} \\[20mm]
\textbf{Abstract}\\[7mm]
{\parbox{13cm}{\hspace{5mm}
The Bagger-Lambert-Gustavsson theory is coupled to an off-shell 
$D=3$, ${\cal N}=8$ conformal supergravity background. 
Local transformation laws of the conformal supergravity multiplet are 
obtained from those of the $D=4$, ${\cal N}=8$ SO(8) gauged supergravity. 
The complete Lagrangian and local transformation laws of the 
coupled theory are obtained. As an application the supercurrent 
multiplet in a flat background is obtained. 
}}
\end{center}
\vfill
\newpage
\setcounter{section}{0}
\setcounter{equation}{0}
\numberbysection
%
%
\newsection{Introduction}
The purpose of this paper is to study a coupling of the 
Bagger-Lambert-Gustavsson (BLG) theory 
\cite{Bagger:2006sk,Gustavsson:2007vu,Bagger:2007jr} 
to a conformal supergravity background. 
The BLG theory is a three-dimensional field theory which was 
proposed to be related to a world-volume theory of M2-branes. 
(For a review, see \cite{Bagger:2012jb}.) 
It is invariant under ${\cal N}=8$ superconformal transformations 
and is naturally coupled to $D=3$, ${\cal N}=8$ conformal supergravity. 
\par

Our main motivation for studying such a coupling comes from 
the AdS/CFT correspondence \cite{Maldacena:1997re}. 
In the AdS/CFT correspondence superstring/M-theory or its 
low energy effective supergravity theory in $(d+1)$-dimensional 
anti de Sitter (AdS) spacetime times a compact manifold 
is dual to a conformal field theory 
on $d$-dimensional boundary at infinity. Boundary values of the 
supergravity fields play a role of sources for operators of the 
conformal field theory \cite{Gubser:1998bc,Witten:1998qj}. 
These boundary supergravity fields at infinity form a $d$-dimensional 
conformal supergravity multiplet \cite{Ferrara:1998ej,Liu:1998bu}. 
Local transformations of the ($d+1$)-dimensional supergravity 
induce local transformations of the $d$-dimensional conformal 
supergravity on these fields \cite{Nishimura:1998ud,
Nishimura:1999gg,Nishimura:1999av,Nishimura:2000wj}. 
A typical example of the AdS/CFT correspondence is the 
correspondence between type IIB superstring or supergravity 
on AdS${}_5 \times $ S${}^5$ 
and $D=4$, ${\cal N}=4$ super Yang-Mills theory. 
The latter theory is invariant under ${\cal N}=4$ superconformal 
transformations and can be coupled to $D=4$, 
${\cal N}=4$ conformal supergravity \cite{Bergshoeff:1980is}. 
Similarly, the BLG theory, which is supposed to be dual to 
M-theory on AdS${}_4 \times S^7$, can be coupled to $D=3$, 
${\cal N}=8$ conformal supergravity. 
\par

Apart from a relation to the AdS/CFT correspondence the coupling 
of the BLG theory to a background conformal supergravity is useful 
when studying its field theoretical properties. 
Conformal supergravity fields couple to operators in the 
supercurrent multiplet of the BLG theory such as the energy-momentum 
tensor and the supercurrents. 
One can use the conformal supergravity fields as sources 
to treat these operators systematically. 
The coupled theory can be used also to obtain a field theory with 
a rigid supersymmetry in a curved spacetime \cite{Festuccia:2011ws} 
by choosing a background invariant under supertransformations. 
\par

A coupling of the BLG theory to $D=3$, ${\cal N}=8$ conformal 
supergravity was previously studied in \cite{Gran:2008qx,Gran:2012mg}. 
The conformal supergravity multiplet considered there in the 
component field formulation contains 
a dreibein, eight Rarita-Schwinger fields and SO(8) gauge fields. 
Local symmetry transformation laws of these fields were obtained, 
and an invariant Lagrangian of the coupled theory was constructed. 
The Lagrangian contains the kinetic terms of the conformal 
supergravity multiplet, and commutators of local supertransformations 
close only when their field equations are used. 
In this sense the conformal supergravity multiplet considered in 
\cite{Gran:2008qx,Gran:2012mg} is an on-shell multiplet. 
\par

Since conformal supergravity we consider in this paper is 
a non-dynamical background, the commutator algebra of local 
transformations should close off-shell. Such an off-shell 
$D=3$, ${\cal N}=8$ conformal supergravity multiplet was 
discussed in the superspace formulation in 
\cite{Howe:1995zm,Cederwall:2011pu,Gran:2012mg}. 
The off-shell multiplet contains scalar and spinor fields 
in addition to the fields of the on-shell multiplet. 
Local supertransformation laws of the component fields 
were explicitly given at the linearized level 
in \cite{Bergshoeff:2010ui}. 
In this paper we will 
obtain a nonlinear form of local transformation laws of 
the off-shell multiplet in the component field formulation. 
There are two kinds of conformal supergravity multiplets 
depending on a sign parameter $\eta=\pm 1$ 
appearing in self-duality conditions on the scalar fields. 
Then, we will construct a complete Lagrangian of the BLG theory coupled 
to the off-shell conformal supergravity background in the $\eta=+1$ case. 
In the $\eta=-1$ case we did not succeed in constructing a coupled theory. 
As we will discuss in section 6, the coupled theory studied 
in the previous works \cite{Gran:2008qx,Gran:2012mg} corresponds 
to the $\eta=-1$ case. 
Therefore, our coupled theory for $\eta=+1$ is the case 
not considered previously. 
As an application of the coupled theory we will discuss 
the supercurrent multiplet of the BLG theory in a flat background. 
Knowledge on off-shell conformal supergravities may 
be also useful to construct new supergravity 
theories \cite{Bergshoeff:2010ui}. 
\par

The organization of this paper is as follows. 
In the next section we briefly explain how to find local 
transformations of the off-shell conformal supergravity multiplet 
starting from the $D=4$, ${\cal N}=8$ SO(8) gauged supergravity. 
Explicit forms of the local transformations are given in section 3. 
In section 4 we obtain local symmetry transformations and an 
invariant Lagrangian of the BLG theory coupled to the conformal 
supergravity background. In section 5 we study the supercurrent 
multiplet of the BLG theory in a flat background. 
In section 6 we discuss difficulties in the $\eta=-1$ case, 
comparisons with the previous works, and truncations of the 
${\cal N}=8$ conformal supergravity multiplet to lower ${\cal N}$. 
In Appendix A we explain our notations and conventions. 
In Appendix B we collect identities of SO(8) matrices, 
which are useful for calculations in the text. 

\vfil

%
\newsection{Conformal supergravity from gauged supergravity}
In this section we briefly sketch how to derive local transformation 
laws of the $D=3$, ${\cal N}=8$ conformal supergravity from those of 
the $D=4$, ${\cal N}=8$ SO(8) gauged supergravity, 
which has a four-dimensional AdS solution. 
Our approach is the same as those in \cite{Nishimura:1998ud,
Nishimura:1999gg,Nishimura:1999av,Nishimura:2000wj} 
and is based on an idea in the AdS/CFT correspondence 
\cite{Maldacena:1997re,Gubser:1998bc,Witten:1998qj}. 
We study boundary behaviors of the $D=4$ gauged supergravity 
at infinity of four-dimensional spacetime. 
Boundary values of the fields form a $D=3$ conformal 
supergravity multiplet. 
We obtain local transformations on these three-dimensional fields 
induced from the four-dimensional local transformations. 
The resulting local transformation laws of the $D=3$, ${\cal N}=8$ 
conformal supergravity are given 
in the next section in a self-contained way. 
\par

The field content of the $D=4$, ${\cal N}=8$ SO(8) gauged 
supergravity \cite{deWit:1982ig} is a vierbein $e_M{}^A(x)$, 
8 Majorana Rarita-Schwinger fields $\psi_M^\alpha(x)$, 
28 SO(8) gauge fields $B_M^{\alpha\beta}(x) = -B_M^{\beta\alpha}(x)$, 
56 Majorana spinor fields 
$\lambda^{\alpha\beta\gamma}(x) = \lambda^{[\alpha\beta\gamma]}(x)$ 
and 70 real scalar fields. Here, $M, N, \cdots = 0,1,2,3$ and 
$A,B, \cdots = 0,1,2,3$ are four-dimensional world and local Lorentz 
indices respectively, and 
$\alpha, \beta, \gamma, \cdots = 1, 2, \cdots, 8$ are SO(8) 
spinor indices of negative chirality. 
(See Appendix A for our notations and conventions.)
The scalar fields take values in the 
coset space E${}_{7(+7)}$/SU(8) \cite{Cremmer:1979up} 
and can be represented by a $56 \times 56$ matrix 
\begin{equation}
V(x) = \exp \left( 
\begin{array}{cc}
0 & \frac{1}{\sqrt{2}} \phi(x) \\
\frac{1}{\sqrt{2}} \phi(x)^* & 0
\end{array}
\right),
\label{expphi}
\end{equation}
where $\phi(x)$ is a complex $28 \times 28$ matrix. 
Components of $\phi(x)$ are denoted as 
$\phi^{\alpha\beta\gamma\delta}(x)$, where the antisymmetrized 
pairs of indices $[\alpha\beta]$ and $[\gamma\delta]$ 
label rows and columns, respectively. 
$\phi^{\alpha\beta\gamma\delta}(x)$ is totally antisymmetric 
in four indices and satisfies a self-duality condition 
\begin{equation}
\phi^{\alpha_1\cdots\alpha_4} 
= \frac{1}{4!} \eta \epsilon^{\alpha_1\cdots\alpha_8} 
(\phi^{\alpha_5\cdots\alpha_8})^*, 
\end{equation}
where $\epsilon^{\alpha_1\cdots\alpha_8}$ is the totally 
antisymmetric symbol with $\epsilon^{12345678} = +1$ and 
$\eta$ is a parameter taking a value $+1$ or $-1$. 
Local symmetries of this theory are those under general coordinate 
transformations, local Lorentz transformations, SO(8) gauge 
transformations and local supertransformations. 
\par

Field equations of this theory have a solution for which 
the metric is given by AdS spacetime and other fields are zero. 
The four-dimensional AdS spacetime is locally represented as 
a region $r \equiv x^3 > 0$ in $\mathbb{R}^4$ with coordinates $x^M$. 
The boundary at infinity of the AdS spacetime corresponds 
to a hyperplane $r = 0$. The AdS metric is given by 
\begin{equation}
dx^M dx^N \bar{g}_{MN} 
= \frac{1}{(2gr)^2} \left( 
dr^2 + dx^\mu dx^\nu \eta_{\mu\nu} \right), 
\label{background}
\end{equation}
where $\mu, \nu = 0,1,2$ are three-dimensional world indices 
and $g$ is the SO(8) gauge coupling constant. 
We will consider fluctuations of the fields around this solution. 
\par

To proceed we partially fix a gauge for the local symmetries. 
We choose gauge fixing conditions as 
\begin{equation}
e_r{}^3 = \frac{1}{2gr}, \qquad
e_r{}^a = 0, \qquad
e_\mu{}^3 = 0, \qquad
\psi_r = 0, \qquad
B_r^{\alpha\beta} = 0, 
\label{gcond}
\end{equation}
where $a, b, \cdots = 0, 1, 2$ are three-dimensional 
local Lorentz indices. 
By solving linearized field equations we can find asymptotic 
behaviors of the fields near the boundary $r = 0$. 
The boundary conditions will be chosen later such that 
they are consistent with these boundary behaviors. 
We assume that $e_\mu{}^a$ behaves as $r^{-1}$ 
just as in the background (\ref{background}). 
Boundary behaviors of other fields are determined by their 
linearized field equations around the background. 
For each field there are two independent solutions which 
have different boundary behaviors. For instance, the vector fields 
$B_\mu^{\alpha\beta}$ have two solutions which behave as 
$r^0$ and $r^1$ for $r \rightarrow 0$. 
The solution regular in the bulk is a linear combination 
of these two solutions and its boundary behavior is dominated 
by a solution which becomes larger near the boundary, i.e., 
$B_\mu^{\alpha\beta} \sim r^0$ in this case. 
To solve the fermionic field equations it is convenient to 
introduce projection operators $\frac{1}{2}(1 \pm \gamma^3)$, 
where $\gamma^3$ is a four-dimensional gamma matrix in the $A=3$ 
direction, and 
define $\psi_{\mu\pm} = \frac{1}{2}(1 \pm \gamma^3)\psi_\mu$, 
$\lambda_\pm = \frac{1}{2}(1 \pm \gamma^3) \lambda$. 
\par

Knowing the boundary behaviors of the fields we impose Dirichlet 
type boundary conditions for $r \rightarrow 0$ as 
\begin{eqnarray}
e_\mu{}^a(x,r) \A\sim\A (2gr)^{-1} e_{0\mu}{}^a(x), \quad
\psi_{0\mu-}(x,r) \sim (2gr)^{-\frac{1}{2}} \psi_{0\mu-}(x), \nonu
B_\mu^{\alpha\beta}(x,r) \A\sim\A g^{-1} B_{0\mu}^{\alpha\beta}(x), \quad
\lambda_-(x,r) \sim (2r)^{\frac{3}{2}} g^{\frac{1}{2}} 
\lambda_{0-}(x), \nonu
\phi^{\alpha\beta\gamma\delta}(x,r) 
\A\sim\A (2r)^2 D_0^{\alpha\beta\gamma\delta} (x)
+ 2ir E_0^{\alpha\beta\gamma\delta}(x), 
\label{bbfield}
\end{eqnarray}
where $\Phi_0  = (e_{0\mu}{}^a, \psi_{0\mu-}, B_{0\mu}, 
\lambda_{0-}, D_0, E_0)$ are arbitrary fixed functions on the boundary. 
The real scalar fields $D_0^{\alpha\beta\gamma\delta}$, 
$E_0^{\alpha\beta\gamma\delta}$ are totally antisymmetric 
in four indices and satisfy (anti) self-duality conditions 
\begin{eqnarray}
D_0^{\alpha_1\cdots\alpha_4} 
\A=\A \frac{1}{4!}\eta 
\epsilon^{\alpha_1\cdots\alpha_8}
D_0^{\alpha_5\cdots\alpha_8}, \nonu
E_0^{\alpha_1\cdots\alpha_4} 
\A=\A - \frac{1}{4!}\eta \epsilon^{\alpha_1\cdots\alpha_8}
E_0^{\alpha_5\cdots\alpha_8}. 
\label{scalarduality}
\end{eqnarray}
Note that we imposed the boundary conditions on only half 
of components of $\psi_\mu$ and $\lambda$ 
since their field equations are first order. 
Other components of these fields behave as 
$\psi_{\mu+}(x,r) \sim (2r)^{\frac{1}{2}} g^{-\frac{1}{2}} \psi_{0\mu+}(x)$, 
$\lambda_+(x,r) \sim (2r)^{\frac{5}{2}} g^{\frac{1}{2}} \lambda_{0+}(x)$. 
By the field equations $\psi_{0\mu+}$, $\lambda_{0+}$ can 
be expressed in terms of $\Phi_0$ (see (\ref{definition1})) 
and therefore cannot be chosen arbitrarily. 
The fields $\Phi_0$ coincide with the field content of the off-shell 
$D=3$, ${\cal N}=8$ conformal supergravity 
\cite{Howe:1995zm, Bergshoeff:2010ui}. 
\par

We now study how the fields $\Phi_0$ transform 
under residual symmetry transformations after the gauge fixing. 
We first obtain parameters of residual symmetries for general 
coordinate, local Lorentz, SO(8) gauge and local 
super transformations, which 
preserve the gauge conditions (\ref{gcond}). 
Parameters of these transformations are denoted as $\xi^M$, 
$\lambda^{AB}=-\lambda^{BA}$, 
$\zeta^{\alpha\beta} = - \zeta^{\beta\alpha}$ and $\epsilon$, 
respectively. By solving differential equations on these parameters 
obtained by taking variations of the gauge conditions (\ref{gcond}) 
under the local transformations we find the parameters of the 
residual symmetry transformations near the boundary $r = 0$ as 
\begin{eqnarray}
\xi^\mu(x,r) \A = \A \xi_0^\mu(x) + {\cal O}(r^2), \nonu
\xi^r(x,r) \A = \A -  r \Lambda_0(x), \nonu
\lambda_{ab}(x,r) \A=\A \lambda_{0ab}(x) + {\cal O}(r^2), \nonu
\lambda_{a3}(x,r) \A=\A {\cal O}(r), \nonu
\zeta^{\alpha\beta}(x,r) \A=\A \zeta_0{}^{\alpha\beta}(x) 
+ {\cal O}(r^2), \nonu
\epsilon_-(x,r) \A=\A (2gr)^{-\frac{1}{2}} \left[ 
\epsilon_{0-}(x) + {\cal O}(r^2) \right], \nonu
\epsilon_+(x,r) \A=\A - (2g)^{-\frac{1}{2}} r^{\frac{1}{2}} 
\left[ \eta_{0+}(x) + {\cal O}(r^2) \right], 
\label{bbparam}
\end{eqnarray}
where $\xi_0^\mu$, $\Lambda_0$, $\lambda_{0ab}$, 
$\zeta_0^{\alpha\beta}$, $\epsilon_{0-}$ and $\eta_{0+}$ 
are arbitrary functions of $x^\mu$ ($\mu=0, 1, 2$). 
Order ${\cal O}(r)$ and ${\cal O}(r^2)$ terms are non-local 
functionals of these functions and the fields $\Phi_0$. 
The transformations with the parameters $\xi_0^\mu$, $\lambda_{0ab}$ 
and $\epsilon_{0-}$ act on the boundary fields $\Phi_0$ as 
three-dimensional general coordinate, local Lorentz and 
local super transformations. 
The transformation with the parameter $\Lambda_0$ acts on the 
boundary fields as Weyl transformation. 
When a four-dimensional field behaves as 
$\varphi(x,r) \sim r^{-w} \varphi_{0}(x)$ for $r \rightarrow 0$ 
the transformation is 
$\delta \varphi_0(x) = w \Lambda_0(x) \varphi_0(x)$. 
The transformation with the parameter $\eta_0$ acts on the 
boundary fields as super Weyl transformation. 
All the transformations of $\Phi_0$ become independent of the SO(8) 
gauge coupling constant $g$ when we scale the fields and 
the transformation parameters 
by $g$ as in (\ref{bbfield}) and (\ref{bbparam}). 
Explicit forms of the transformations obtained in this way 
will be given in the next section. 
\par

The supertransformations of the $D=4$ gauged supergravity 
are non-polynomial in the scalar fields $\phi$ as can be seen 
in the expression (\ref{expphi}). However, in the limit 
$r \rightarrow 0$ they become polynomials in $D_0$ and $E_0$. 
Similarly, although the supertransformations of the fermionic fields 
in the $D=4$ gauged supergravity contain terms quadratic in the 
fermionic fields, most of them go to zero in the limit $r \rightarrow 0$. 
Only remaining such higher Fermi terms are those appearing in 
supercovariant derivatives.  
\par

In the following we will suppress the subscript $0$ for the 
boundary fields and the transformation parameters. 
Four-component spinors of SO(1,3) with a definite $\gamma^3$ 
eigenvalue will be expressed by two-component spinors of SO(1,2), 
and the subscripts $\pm$ of $\psi_{0\mu-}$, $\lambda_{0-}$, 
$\epsilon_{0-}$, $\eta_{0+}$ will be suppressed. 


%
\newsection{\boldmath{$D=3$}, \boldmath{${\cal N}=8$} 
conformal supergravity}
Let us summarize the results on the $D=3$, ${\cal N}=8$ conformal 
supergravity obtained in the last section. 
The field content is, as given in Table 1, a dreibein $e_\mu{}^a(x)$, 
8 Majorana Rarita-Schwinger fields $\psi_\mu^\alpha(x)$, 
28 SO(8) gauge fields $B_\mu^{\alpha\beta}(x)$, 
56 Majorana spinor fields $\lambda^{\alpha\beta\gamma}(x)$, 
70 real scalar fields $D^{\alpha\beta\gamma\delta}(x)$, 
$E^{\alpha\beta\gamma\delta}(x)$, where 
$\alpha, \beta, \cdots = 1,2,\cdots,8$ are SO(8) negative chirality 
spinor indices. 
The fields $B_\mu^{\alpha\beta}(x)$, $\lambda^{\alpha\beta\gamma}(x)$, 
$D^{\alpha\beta\gamma\delta}(x)$, $E^{\alpha\beta\gamma\delta}(x)$ 
are totally antisymmetric in these indices. 
The scalar fields also satisfy (anti) self-duality conditions 
\begin{eqnarray}
D^{\alpha_1\cdots\alpha_4} 
\A=\A \frac{1}{4!}\eta 
\epsilon^{\alpha_1\cdots\alpha_8}
D^{\alpha_5\cdots\alpha_8}, \nonu
E^{\alpha_1\cdots\alpha_4} 
\A=\A - \frac{1}{4!}\eta \epsilon^{\alpha_1\cdots\alpha_8}
E^{\alpha_5\cdots\alpha_8}, 
\label{scalarduality2}
\end{eqnarray}
where $\eta$ is a parameter taking a value $+1$ or $-1$. 
\par

\begin{table}[t]
\begin{center}
\begin{tabular}{|c|c|c|c|c|c|c|}
\hline
Field & $e_\mu{}^a$ & $\psi_\mu^\alpha$ & $B_\mu^{\alpha\beta}$ & 
$\lambda^{\alpha\beta\gamma}$ & $D^{\alpha\beta\gamma\delta}$ & 
$E^{\alpha\beta\gamma\delta}$ \\ \hline
Weyl weight & $1$ & $\frac{1}{2}$ & 0 & $-\frac{3}{2}$ & $-2$ & $-1$ \\
\hline 
SO(8) representation & {\bf 1} & ${\bf 8}_s$ & {\bf 28} & ${\bf 56}_s$ 
& ${\bf 35}_v$ or ${\bf 35}_c$
& ${\bf 35}_c$ or ${\bf 35}_v$ \\ \hline
\end{tabular}
\caption{The field content of the conformal supergravity multiplet.
SO(8) representations of $D^{\alpha\beta\gamma\delta}$ 
and $E^{\alpha\beta\gamma\delta}$ are the first ones for $\eta=+1$ 
and the second ones for $\eta=-1$.}
\end{center}
\end{table}

Local symmetry transformations of these fields are 
general coordinate transformation $\delta_G$ with a  
parameter $\xi^\mu(x)$, local Lorentz transformation $\delta_L$ 
with a parameter $\lambda_{ab}(x) = - \lambda_{ba}(x)$, 
Weyl transformation $\delta_W$ with a parameter $\Lambda(x)$, 
local SO(8) transformation $\delta_g$ with a parameter 
$\zeta^{\alpha\beta}(x) = - \zeta^{\beta\alpha}(x)$, 
local supertransformation $\delta_Q$ with a Majorana spinor parameter 
$\epsilon^\alpha(x)$ and super Weyl transformation $\delta_S$ 
with a Majorana spinor parameter $\eta^\alpha(x)$. 
Weyl weights and SO(8) representations of the fields are given in 
Table 1. (See Tables 36, 37 of \cite{Slansky:1981yr} for naming 
of SO(8) representations.)
Bosonic transformation laws other than the Weyl transformation
are obvious from the index structure of the fields. 
For instance, the bosonic transformations of the
Rarita-Schwinger fields with Weyl weight $\frac{1}{2}$ are 
\begin{equation}
(\delta_G + \delta_L + \delta_W + \delta_g) \psi_\mu^\alpha
= \xi^\nu \partial_\nu \psi_\mu^\alpha 
+ \partial_\mu \xi^\nu \psi_\nu^\alpha 
- \frac{1}{4} \lambda_{ab} \gamma^{ab} \psi_\mu^\alpha 
+ \frac{1}{2} \Lambda \psi_\mu^\alpha 
- \zeta^{\alpha\beta} \psi_\mu^\beta. 
\end{equation}
The fermionic transformations $\delta_Q$ and $\delta_S$ are given by 
\begin{eqnarray}
\delta_Q e_\mu{}^a 
\A=\A \frac{1}{4} \bar{\epsilon}^\alpha \gamma^a \psi_\mu^\alpha, \qquad
\delta_Q \psi_\mu^\alpha = D_\mu \epsilon^\alpha, \nonu
\delta_Q B_\mu^{\alpha\beta} 
\A=\A - \bar{\epsilon}^{[\alpha} \psi_{\mu+}^{\beta]} 
+ \frac{1}{2\sqrt{2}} \bar{\epsilon}^\gamma \gamma_\mu 
\lambda^{\alpha\beta\gamma} 
- \frac{1}{2\sqrt{2}} \bar{\epsilon}^\gamma \psi_\mu^\delta 
E^{\alpha\beta\gamma\delta}, \nonu
\delta_Q \lambda^{\alpha\beta\gamma} 
\A=\A - \frac{3}{4\sqrt{2}} \gamma^{\mu\nu} \epsilon^{[\alpha} 
\hat{G}_{\mu\nu}^{\beta\gamma]}
+ \epsilon^\delta D^{\alpha\beta\gamma\delta} 
- \frac{1}{2} \gamma^\mu \epsilon^\delta 
\hat{D}_\mu E^{\alpha\beta\gamma\delta} \nonu
\A\A \mbox{} - \frac{3}{4\sqrt{2}} \epsilon^\delta 
E^{\epsilon\eta[\alpha\beta} E^{\gamma\delta]\epsilon\eta}, \nonu
\delta_Q D^{\alpha_1\cdots\alpha_4} 
\A=\A \mbox{} \bar{\epsilon}^{[\alpha_1} \lambda_+^{\alpha_2\alpha_3\alpha_4]} 
+ \frac{1}{4!} \eta \epsilon^{\alpha_1\cdots\alpha_8} 
\bar{\epsilon}^{\alpha_5} \lambda_+^{\alpha_6\alpha_7\alpha_8}, \nonu
\delta_Q E^{\alpha_1\cdots\alpha_4} 
\A=\A \bar{\epsilon}^{[\alpha_1} \lambda^{\alpha_2\alpha_3\alpha_4]} 
- \frac{1}{4!} \eta \epsilon^{\alpha_1\cdots\alpha_8} 
\bar{\epsilon}^{\alpha_5} \lambda^{\alpha_6\alpha_7\alpha_8} 
\label{supertransformation}
\end{eqnarray}
and 
\begin{eqnarray}
\delta_S e_\mu{}^a \A=\A 0, \qquad
\delta_S \psi_\mu^\alpha = \gamma_\mu \eta^\alpha, \qquad
\delta_S B_\mu^{\alpha\beta} 
= \frac{1}{2} \bar{\eta}^{[\alpha} \psi_\mu^{\beta]}, \nonu
\delta_S \lambda^{\alpha\beta\gamma} 
\A=\A \eta^\delta E^{\alpha\beta\gamma\delta}, \qquad
\delta_S E^{\alpha_1\cdots\alpha_4} = 0, \nonu
\delta_S D^{\alpha_1\cdots\alpha_4} 
\A=\A - \frac{1}{2} \left( \bar{\eta}^{[\alpha_1} 
\lambda^{\alpha_2\alpha_3\alpha_4]} 
+ \frac{1}{4!} \eta \epsilon^{\alpha_1\cdots\alpha_8} 
\bar{\eta}^{[\alpha_5} \lambda^{\alpha_6\alpha_7\alpha_8]}  
\right).
\label{superweyl}
\end{eqnarray}
Here, we have defined 
\begin{eqnarray}
\psi_{\mu+}^\alpha 
\A=\A \frac{1}{4} \gamma^{\rho\sigma} \gamma_\mu 
\psi_{\rho\sigma}^\alpha, \qquad
\psi_{\mu\nu}^\alpha = D_{[\mu} \psi_{\nu]}^\alpha, \nonu
\lambda_+^{\alpha\beta\gamma} 
\A=\A - \frac{1}{2} \gamma^\mu \hat{D}_\mu \lambda^{\alpha\beta\gamma} 
- \gamma^\mu \psi_{\mu+}^\delta E^{\alpha\beta\gamma\delta} 
+ \frac{3}{2\sqrt{2}} E^{\delta\epsilon[\alpha\beta} 
\lambda^{\gamma]\delta\epsilon}, \nonu
\hat{G}_{\mu\nu}^{\alpha\beta} 
\A=\A G_{\mu\nu}^{\alpha\beta} 
+ 2 \bar{\psi}_{[\mu}^{[\alpha} \psi_{\nu]+}^{\beta]} 
- \frac{1}{\sqrt{2}} \bar{\psi}_{[\mu}^{\gamma} \gamma_{\nu]} 
\lambda^{\alpha\beta\gamma}
+ \frac{1}{2\sqrt{2}} \bar{\psi}_{[\mu}^\gamma \psi_{\nu]}^\delta 
E^{\alpha\beta\gamma\delta}, \nonu
\hat{D}_\mu E^{\alpha_1\cdots\alpha_4} 
\A=\A D_\mu E^{\alpha_1\cdots\alpha_4} 
- \bar{\psi}_\mu^{[\alpha_1} \lambda^{\alpha_2\alpha_3\alpha_4]} 
+ \frac{1}{4!} \eta \epsilon^{\alpha_1\cdots\alpha_8} 
\bar{\psi}_\mu^{\alpha_5} \lambda^{\alpha_6\alpha_7\alpha_8}, \nonu
\hat{D}_\mu \lambda^{\alpha\beta\gamma} \A=\A 
D_\mu \lambda^{\alpha\beta\gamma} 
+ \frac{3}{4\sqrt{2}} \, \gamma^{\rho\sigma} \psi_\mu^{[\alpha}
\hat{G}_{\rho\sigma}^{\beta\gamma]} 
- \psi_\mu^\delta D^{\alpha\beta\gamma\delta} 
+ \frac{1}{2} \gamma^\rho \psi_\mu^\delta \hat{D}_\rho 
E^{\alpha\beta\gamma\delta} \nonu
\A\A \mbox{} + \frac{3}{4\sqrt{2}} \, \psi_\mu^\delta 
E^{\epsilon\eta[\alpha\beta} E^{\gamma\delta]\epsilon\eta}. 
\label{definition1}
\end{eqnarray}
The covariant derivative $D_\mu$ contains the spin connection and 
the SO(8) gauge fields and is given by, e.g., for $\epsilon^\alpha$
\begin{equation}
D_\mu \epsilon^\alpha = \left( \partial_\mu 
+ \frac{1}{4} \hat{\omega}_{\mu ab} \gamma^{ab} \right) \epsilon^\alpha
+ B_\mu^{\alpha\beta} \epsilon^\beta. 
\end{equation}
The spin connection $\hat{\omega}_{\mu ab}$ satisfies 
the torsion condition 
\begin{equation}
D_\mu e_\nu{}^a - D_\nu e_\mu{}^a 
= \frac{1}{4} \bar{\psi}_\mu \gamma^a \psi_\nu
\end{equation}
and is given by 
\begin{equation}
\hat{\omega}_{\mu ab} 
= \omega_{\mu ab}(e) + \frac{1}{8} ( \bar{\psi}_a \gamma_\mu \psi_b 
+ \bar{\psi}_\mu \gamma_a \psi_b 
- \bar{\psi}_\mu \gamma_b \psi_a ),  
\label{spinconnection}
\end{equation}
where $\omega_{\mu ab}(e)$ is the spin connection without torsion. 
The Ricci tensor made from the spin connection $\hat{\omega}_{\mu ab}$ 
is not a symmetric tensor but has an antisymmetric part 
\begin{equation}
R_{[\mu\nu]} 
= - \frac{3}{4} \bar{\psi}_{[\rho} \gamma^\rho \psi_{\mu\nu]}. 
\end{equation}
$\hat{D}_\mu$ is the supercovariant derivative, which transforms 
without $\partial_\mu \epsilon$ terms under the local 
supertransformation. $\hat{D}_\mu E^{\alpha\beta\gamma\delta}$ 
satisfies the same self-duality condition as $E^{\alpha\beta\gamma\delta}$ 
in (\ref{scalarduality2}). 
The field strength of the SO(8) gauge fields is 
\begin{equation}
G_{\mu\nu}^{\alpha\beta} 
= \partial_\mu B_\nu^{\alpha\beta} 
- \partial_\nu B_\mu^{\alpha\beta}
+ B_\mu^{\alpha\gamma} B_\nu^{\gamma\beta} 
- B_\nu^{\alpha\gamma} B_\mu^{\gamma\beta}, 
\end{equation}
and $\hat{G}_{\mu\nu}^{\alpha\beta}$ is its supercovariantization. 
By linearization the transformations (\ref{supertransformation}), 
(\ref{superweyl}) reduce to the transformations 
in \cite{Howe:1995zm, Bergshoeff:2010ui}. 
\par

The above local transformations satisfy a commutator algebra
which closes off-shell, i.e., without using any constraints 
on the fields such as field equations. 
This is guaranteed from the fact that the local transformations of 
the $D=4$ $SO(8)$ gauged supergravity satisfy a closed commutator algebra 
on-shell and the fact that the field equations of the $D=4$ theory 
do not impose any constraint on the boundary values $\Phi_0  
= (e_{0\mu}{}^a, \psi_{0\mu-}, B_{0\mu}, \lambda_{0-}, D_0, E_0)$, 
which become the fields of the $D=3$ conformal supergravity 
multiplet as discussed in the previous section. 
Explicit forms of the commutators will be given below. 
\par

So far we have used only negative chirality spinor indices 
$\alpha, \beta, \cdots = 1,2,\cdots,8$ of SO(8). On the other hand 
the BLG theory contains fields with vector indices 
$I, J, \cdots = 1,2,\cdots,8$ and positive chirality spinor indices 
$\dot{\alpha}, \dot{\beta}, \cdots = 1,2,\cdots,8$ of SO(8). 
To couple the conformal supergravity to the BLG theory 
we need the Clebsch-Gordan coefficients which connect these three 
kinds of indices of SO(8). 
They are provided by the $16 \times 16$ gamma matrices of SO(8)
\begin{equation}
\Gamma^I = \left( 
\begin{array}{cc}
0 & \Sigma^I \\
\bar{\Sigma}^I & 0
\end{array}
\right), 
\label{so8gamma}
\end{equation}
where $\Sigma^I$ and $\bar{\Sigma}^I = (\Sigma^I)^T$ are $8 \times 8$ 
matrices with components $(\Sigma^I)^{\alpha\dot{\beta}}$ and 
$(\bar{\Sigma}^I)^{\dot{\alpha}\beta}$. 
These $\Sigma$-matrices satisfy identities given in Appendix B. 
\par

To couple the conformal supergravity to the BLG theory 
it is convenient to define fields with vector and positive chirality 
spinor indices instead of negative chirality spinor indices. 
For that purpose we introduce combinations of 
$\Sigma$-matrices\footnote{These were used in \cite{Green:1983hw} 
in a study of the superstring field theory in the light-cone gauge. 
See Appendix A of \cite{Green:1983hw} for further identities 
which they satisfy.}  
\begin{eqnarray}
t_{IJ}^{\alpha\beta\gamma\delta} 
\A=\A (\Sigma_{KI})^{[\alpha\beta} (\Sigma_{KJ})^{\gamma\delta]}, \nonu
t_{IJKL}^{\alpha\beta\gamma\delta} 
\A=\A (\Sigma_{[IJ})^{[\alpha\beta} (\Sigma_{KL]})^{\gamma\delta]}, \nonu
u_{I\dot{\alpha}}^{\alpha\beta\gamma} 
\A=\A - (\Sigma_{IJ})^{[\alpha\beta} (\Sigma_J)^{\gamma]\dot{\alpha}}, 
\end{eqnarray}
where $\Sigma^{IJ} = \Sigma^{[I} \bar{\Sigma}^{J]}$, and 
upper and lower indices are not distinguished. They satisfy
\begin{eqnarray}
t_{IJ}^{\alpha_1\cdots\alpha_4} 
\A=\A \frac{1}{4!} \epsilon^{\alpha_1\cdots\alpha_8} 
t_{IJ}^{\alpha_5\cdots\alpha_8}, \quad
t_{IJKL}^{\alpha_1\cdots\alpha_4} 
= - \frac{1}{4!} \epsilon^{\alpha_1\cdots\alpha_8} 
t_{IJKL}^{\alpha_5\cdots\alpha_8}, \nonu
t_{IJ}^{\alpha\beta\gamma\delta} 
= t_{JI}^{\alpha\beta\gamma\delta}, \A\A \quad\!\!\!\!\!\!\!
t_{II}^{\alpha\beta\gamma\delta} = 0, \quad
t_{I_1\cdots I_4}^{\alpha\beta\gamma\delta} 
= \frac{1}{4!} \epsilon^{I_1\cdots I_8}
t_{I_5\cdots I_8}^{\alpha\beta\gamma\delta}, \quad
(\Sigma^I)^{\alpha\dot{\beta}} u_{I\dot{\beta}}^{\beta\gamma\delta} 
= 0. 
\end{eqnarray}
Note that $t_{IJ}^{\alpha_1\cdots\alpha_4}$ is self-dual and 
$t_{IJKL}^{\alpha_1\cdots\alpha_4}$ is anti self-dual 
in the four spinor indices. 
We can define new SO(8) gauge fields $B_\mu^{IJ}$ and spinor fields 
$\lambda^{I\dot{\alpha}}$ by 
\begin{eqnarray}
B_{\mu}^{IJ} 
\A=\A \frac{1}{4} (\Sigma^{IJ})^{\alpha\beta} B_{\mu}^{\alpha\beta}, \qquad
B_\mu^{\alpha\beta} 
= \frac{1}{4} (\Sigma^{IJ})^{\alpha\beta} B_{\mu}^{IJ}, \nonu
\lambda^{I\dot{\alpha}} \A=\A \frac{1}{48} \, 
u_{I\dot{\alpha}}^{\alpha\beta\gamma} \lambda^{\alpha\beta\gamma}, 
\ \qquad
\lambda^{\alpha\beta\gamma} = u_{I\dot{\alpha}}^{\alpha\beta\gamma} 
\lambda^{I\dot{\alpha}},  
\label{newblambda}
\end{eqnarray}
which satisfy 
\begin{equation}
B_{\mu}^{IJ} = - B_\mu^{JI}, \qquad
(\Sigma^I)^{\alpha\dot{\beta}} \lambda^{I\dot{\beta}} = 0. 
\label{newblambda2}
\end{equation}
The field strength of the new gauge field is 
\begin{equation}
G_{\mu\nu}^{IJ} 
=  \partial_\mu B_\nu^{IJ} 
- \partial_\nu B_\mu^{IJ} 
+ B_\mu^{IK} B_\nu^{KJ} - B_\nu^{IK} B_\mu^{KJ}. 
\end{equation}
Definitions of new scalar fields depend on the self-duality parameter 
$\eta$ in (\ref{scalarduality2}). 
When the parameter is $\eta=+1$, $D^{\alpha\beta\gamma\delta}$ is 
self-dual and $E^{\alpha\beta\gamma\delta}$ is anti self-dual. 
Then, we can define the scalar fields $D^{IJ}$ and $E^{IJKL}$ by 
\begin{eqnarray}
D^{IJ} \A=\A \frac{1}{4\cdot4!} \, t_{IJ}^{\alpha\beta\gamma\delta} 
D^{\alpha\beta\gamma\delta}, \qquad
D^{\alpha\beta\gamma\delta} = 
\frac{1}{4} t_{IJ}^{\alpha\beta\gamma\delta} D^{IJ}, \nonu
E^{IJKL} \A=\A \frac{1}{256} \, t_{IJKL}^{\alpha\beta\gamma\delta} 
E^{\alpha\beta\gamma\delta}, \ \qquad
E^{\alpha\beta\gamma\delta} = t_{IJKL}^{\alpha\beta\gamma\delta} 
E^{IJKL}, 
\label{newde}
\end{eqnarray}
which satisfy 
\begin{equation}
D^{IJ} = D^{JI}, \quad
D^{II}= 0, \quad
E^{IJKL} = \frac{1}{4!} \epsilon^{IJKLMNPQ} E^{MNPQ}. 
\label{newde2}
\end{equation}
(One can further define $E^{\dot{\alpha}\dot{\beta}}
= (\bar{\Sigma}^{IJKL})^{\dot{\alpha}\dot{\beta}} E^{IJKL}$, 
which is symmetric and traceless in $(\dot{\alpha}\dot{\beta})$, 
but we use $E^{IJKL}$ in the following.)
On the other hand, when the parameter is $\eta=-1$, we can define 
$D^{IJKL}$ and $E^{IJ}$ as in (\ref{newde}), (\ref{newde2}) 
with $D$ and $E$ interchanged. 
Note that the two cases $\eta=\pm 1$ are not equivalent 
since $D$ and $E$ have different Weyl weights. 
In the following we rewrite the fermionic transformations in 
terms of the new fields.

\subsection{$\eta=+1$}
In this case the fermionic transformations 
$\delta_Q$ and $\delta_S$ in (\ref{supertransformation}) 
and (\ref{superweyl}) become 
\begin{eqnarray}
\delta_Q e_\mu{}^a 
\A=\A \frac{1}{4} \bar{\epsilon} \gamma^a \psi_\mu, \qquad
\delta_Q \psi_\mu = D_\mu \epsilon, \nonu
\delta_Q B_\mu^{IJ}
\A=\A - \frac{1}{4} \bar{\epsilon} \Sigma^{IJ} \psi_{\mu+}
+ \sqrt{2} \bar{\epsilon} \gamma_\mu \Sigma^{[I} \lambda^{J]} 
- \sqrt{2} \bar{\epsilon} \Sigma^{KL} \psi_\mu E^{IJKL}, \nonu
\delta_Q \lambda^{I}
\A=\A - \frac{1}{128\sqrt{2}} \gamma^{\mu\nu} \left( 
\bar{\Sigma}^{IKL} - 6 \delta^{IK} \bar{\Sigma}^L \right) \epsilon
\hat{G}_{\mu\nu}^{KL}
+ \frac{1}{4} \bar{\Sigma}^J \epsilon D^{IJ} \nonu
\A\A \mbox{} + \frac{1}{6} \gamma^\mu \bar{\Sigma}^{JKL} \epsilon 
\hat{D}_\mu E^{IJKL} 
+ 2 \sqrt{2} \, \bar{\Sigma}^{JKL} \epsilon E^{MN[IJ} E^{KL]MN}, \nonu
\delta_Q D^{IJ}
\A=\A - \bar{\epsilon} \Sigma^{(I}\lambda_+^{J)}, \qquad
\delta_Q E^{IJKL}
= \frac{1}{8} \bar{\epsilon} \Sigma^{[IJK} \lambda^{L]}
\label{super1}
\end{eqnarray}
and 
\begin{eqnarray}
\delta_S e_\mu{}^a \A=\A 0, \quad
\delta_S \psi_\mu = \gamma_\mu \eta, \quad
\delta_S B_\mu^{IJ} 
= \frac{1}{8} \bar{\eta} \Sigma^{IJ} \psi_\mu, \nonu
\delta_S \lambda^{I}
\A=\A - \frac{1}{3} \bar{\Sigma}^{JKL} \eta E^{IJKL}, \quad
\delta_S D^{IJ} 
= \frac{1}{2} \bar{\eta} \Sigma^{(I} \lambda^{J)}, \quad
\delta_S E^{IJKL} = 0,  
\end{eqnarray}
where 
\begin{equation}
\lambda_+^{I} = - \frac{1}{2} \gamma^\mu \hat{D}_\mu \lambda^{I} 
+ \frac{1}{3} \gamma^\mu \bar{\Sigma}^{JKL} 
\psi_{\mu+} E^{IJKL} 
+ \frac{1}{\sqrt{2}} \left( \bar{\Sigma}^{IJKL} 
-5 \delta^{IJ} \bar{\Sigma}^{KL} \right) E^{JKLM} \lambda^M. 
\label{lambda+}
\end{equation}
We have suppressed SO(8) spinor indices $\alpha, \dot{\alpha}, \cdots$. 
Explicit forms of the supercovariant quantities $\hat{G}_{\mu\nu}^{IJ}$, 
$\hat{D}_\mu E^{IJKL}$, $\hat{D}_\mu \lambda^I$ 
in terms of the new fields can be found from 
(\ref{definition1}) or more easily from the supertransformations 
(\ref{super1}). $\lambda_+^I$ in (\ref{lambda+}) is related to 
$\lambda_+^{\alpha\beta\gamma}$ in (\ref{definition1}) 
as $\lambda_+^{I\dot{\alpha}} 
= \frac{1}{48} u^{\alpha\beta\gamma}_{I\dot{\alpha}} 
\lambda_+^{\alpha\beta\gamma}$ and 
satisfies $\Sigma^I \lambda_+^I = 0$. 
\par

As explained above the commutator algebra of the local 
transformations closes off-shell. 
Indeed, by lengthy but straightforward calculations using SO(8) 
identities in Appendix B we found the off-shell commutation 
relations as 
\begin{eqnarray}
[ \delta_Q(\epsilon_1), \delta_Q(\epsilon_2) ] 
\A=\A \delta_G(\xi) + \delta_L(\lambda) + \delta_g(\zeta) 
+ \delta_Q(\epsilon') + \delta_S(\eta'), \nonu
[ \delta_Q(\epsilon), \delta_S(\eta) ] 
\A=\A \delta_W(\Lambda) + \delta_L(\lambda') + \delta_g(\zeta') 
+ \delta_S(\eta''), \nonu
[ \delta_S(\eta_1), \delta_S(\eta_2) ] \A=\A 0, 
\label{commutator}
\end{eqnarray}
where the transformation parameters on the right-hand sides are 
\begin{eqnarray}
\xi^\mu \A=\A \frac{1}{4} \bar{\epsilon}_2 \gamma^\mu \epsilon_1, \quad
\lambda_{ab} = - \xi^\mu \hat{\omega}_{\mu ab}, \nonu
\zeta^{IJ} \A=\A - \xi^\mu B_\mu^{IJ}
- \sqrt{2} \bar{\epsilon}_2 \Sigma^{KL} \epsilon_1 E^{IJKL}, \quad
\epsilon' = - \xi^\mu \psi_\mu, \nonu
\eta' \A=\A \frac{1}{32} \xi^\mu ( 14 \gamma^{\rho\sigma} \gamma_\mu 
- \gamma_\mu \gamma^{\rho\sigma} ) \psi_{\rho\sigma} 
- \left( \frac{1}{256} \Sigma^{IJ} \gamma^{\rho\sigma} 
\psi_{\rho\sigma} 
+ \frac{1}{2\sqrt{2}} \Sigma^I \lambda^J \right) 
\bar{\epsilon}_2 \Sigma^{IJ} \epsilon_1 \nonu
\A\A \mbox{} - \frac{1}{256 \cdot 4!} \Sigma^{IJKL} 
( 2 \gamma^{\rho\sigma} \gamma_\mu + \gamma_\mu \gamma^{\rho\sigma})
\psi_{\rho\sigma} \bar{\epsilon}_2 \gamma^\mu 
\Sigma^{IJKL} \epsilon_1, \nonu
\Lambda \A=\A - \frac{1}{4} \bar{\epsilon} \eta, \quad
\lambda'_{ab} = \frac{1}{4} \bar{\epsilon} \gamma_{ab} \eta, \quad
\zeta'^{IJ} = - \frac{1}{8} \bar{\epsilon} \Sigma^{IJ} \eta, \quad
\eta'' = \frac{1}{8} \gamma^\mu \epsilon \bar{\eta} \psi_\mu.
\label{rhsparameters}
\end{eqnarray}
The SO(8) parameter $\zeta^{IJ}$ is related to $\zeta^{\alpha\beta}$ 
as $\zeta^{IJ} = \frac{1}{4} (\Sigma^{IJ})^{\alpha\beta} 
\zeta^{\alpha\beta}$. 
We have explicitly checked these commutation relations on all 
the fields except for $[\delta_Q, \delta_Q] \lambda^I$ and 
$[\delta_Q, \delta_Q] D^{IJ}$, which are the hardest to calculate.

\subsection{$\eta=-1$}
In this case the fermionic transformations 
$\delta_Q$ and $\delta_S$ in (\ref{supertransformation}) 
and (\ref{superweyl}) become 
\begin{eqnarray}
\delta_Q e_\mu{}^a 
\A=\A \frac{1}{4} \bar{\epsilon} \gamma^a \psi_\mu, \qquad
\delta_Q \psi_\mu = D_\mu \epsilon, \nonu
\delta_Q B_\mu^{IJ}
\A=\A - \frac{1}{4} \bar{\epsilon} \Sigma^{IJ} \psi_{\mu+}
+ \sqrt{2} \bar{\epsilon} \gamma_\mu \Sigma^{[I} \lambda^{J]} 
+ \frac{1}{2\sqrt{2}} \bar{\epsilon} \Sigma^{K[I} \psi_\mu E^{J]K}, \nonu
\delta_Q \lambda^{I}
\A=\A - \frac{1}{128\sqrt{2}} \, \gamma^{\mu\nu} \left( 
\bar{\Sigma}^{IKL} - 6 \delta^{IK} \bar{\Sigma}^L \right) \epsilon
\hat{G}_{\mu\nu}^{KL}
- \frac{1}{3} \bar{\Sigma}^{JKL} \epsilon D^{IJKL} \nonu
\A\A \mbox{} - \frac{1}{8} \gamma^\mu \bar{\Sigma}^J \epsilon 
\hat{D}_\mu E^{IJ} 
- \frac{1}{4\sqrt{2}} \, \bar{\Sigma}^J \epsilon \left( 
E^{IK} E^{JK} - \frac{1}{8} \delta^{IJ} E^{KL} E^{KL} \right), \nonu
\delta_Q D^{IJKL}
\A=\A \frac{1}{8} \, \bar{\epsilon} \Sigma^{[IJK} \lambda_+^{L]}, \qquad
\delta_Q E^{IJ}
= - \bar{\epsilon} \Sigma^{(I}\lambda^{J)}
\label{super2}
\end{eqnarray}
and
\begin{eqnarray}
\delta_S e_\mu{}^a \A=\A 0, \quad
\delta_S \psi_\mu = \gamma_\mu \eta, \quad
\delta_S B_\mu^{IJ} 
= \frac{1}{8} \bar{\eta} \Sigma^{IJ} \psi_\mu, \nonu
\delta_S \lambda^{I}
\A=\A \frac{1}{4} \bar{\Sigma}^J \eta E^{IJ}, \quad
\delta_S D^{IJKL} 
= - \frac{1}{16} \bar{\eta} \Sigma^{[IJK} \lambda^{L]}, \quad
\delta_S E^{IJ} = 0,  
\end{eqnarray}
where 
\begin{equation}
\lambda_+^{I} = - \frac{1}{2} \gamma^\mu \hat{D}_\mu \lambda^{I} 
- \frac{1}{4} \gamma^\mu \bar{\Sigma}^J \psi_{\mu+} E^{IJ}
- \frac{1}{4\sqrt{2}} \left( \bar{\Sigma}^{IK} - 7 \delta^{IK} \right) 
\lambda^L E^{KL}.
\end{equation}
The commutators of the fermionic transformations are given by 
(\ref{commutator}). The parameters on the right-hand sides are 
given by (\ref{rhsparameters}) except for 
\begin{equation}
\zeta^{IJ} = - \xi^\mu B_\mu^{IJ}
+ \frac{1}{2\sqrt{2}} \bar{\epsilon}_2 \Sigma^{K[I} 
\epsilon_1 E^{J]K}.
\end{equation}

%
\newsection{The BLG theory coupled to conformal supergravity}
The BLG theory \cite{Bagger:2006sk,Gustavsson:2007vu,Bagger:2007jr} 
is a three-dimensional field theory invariant under ${\cal N}=8$ 
superconformal transformations. 
It is based on an algebraic structure called a 3-algebra 
with a four-index structure constant $f^{klj}{}_i$. 
The field content of the theory is real scalar fields $X_i^I(x)$, 
Majorana spinor fields $\Psi_i^{\dot{\alpha}}(x)$ and 
Chern-Simons gauge fields $\tilde{A}_\mu{}^j{}_i(x) 
= A_{\mu kl}(x) f^{klj}{}_i$ ($A_{\mu kl} = - A_{\mu lk}$) 
as shown in Table 2. 
Here, $I,J,\cdots=1,2,\cdots,8$ and 
$\dot{\alpha},\dot{\beta},\cdots=1,2,\cdots,8$ are 
vector indices and positive chirality spinor indices of SO(8) 
respectively as in the previous section, and 
$i,j,\cdots=1,2,\cdots,n$ are indices of the 3-algebra. 
The 3-algebra indices are raised and lowered by a constant metric 
$h^{ij}$, $h_{ij}$. 
The structure constant of the 3-algebra $f^{ijk}{}_l$ satisfies 
\begin{equation}
f^{ijkl} = f^{[ijkl]}, \qquad
f^{mn[i}{}_p f^{jkl]p} = 0.
\end{equation}
The 3-algebra gauge transformations of the fields are 
\begin{eqnarray}
\delta_{g3} X_i^I \A=\A X_j^I \tilde{\Lambda}^j{}_i, \qquad
\delta_{g3} \Psi_i = \Psi_j \tilde{\Lambda}^j{}_i, \nonu
\delta_{g3} \tilde{A}_\mu{}^j{}_i 
\A=\A D_\mu \tilde{\Lambda}^j{}_i 
= \partial_\mu \tilde{\Lambda}^j{}_i 
- \tilde{\Lambda}^j{}_k \tilde{A}_\mu{}^k{}_i
+ \tilde{A}_\mu{}^j{}_k \tilde{\Lambda}{}^k{}_i,  
\end{eqnarray}
where $\tilde{\Lambda}{}^j{}_i(x) = \Lambda_{kl}(x) f^{klj}{}_i$ 
is a transformation parameter. 
\begin{table}[t]
\begin{center}
\begin{tabular}{|c|c|c|c|}
\hline
Field & $X_i^I$ & $\Psi_i^{\dot{\alpha}}$ & $\tilde{A}_\mu{}^j{}_i$ 
\\ \hline
Weyl weight & $-\frac{1}{2}$ & $-1$ & 0 \\
\hline 
SO(8) representation & ${\bf 8}_v$ & ${\bf 8}_c$ & {\bf 1} \\ \hline
\end{tabular}
\caption{The field content of the BLG multiplet.}
\end{center}
\end{table}

We would like to couple the BLG theory to the off-shell conformal 
supergravity constructed in the previous sections. 
We will only consider the $\eta=+1$ case since we find difficulties 
in coupling the $\eta=-1$ conformal supergravity to the BLG theory 
as will be discussed in section 6. 
By the standard Noether procedure starting from the theory 
in a flat background \cite{Bagger:2006sk,Gustavsson:2007vu,Bagger:2007jr} 
we can obtain local transformation laws of the fields and a Lagrangian 
in the conformal supergravity background. 
\par

First we shall give local transformation laws of the BLG fields. 
Transformations of the conformal supergravity fields remain the 
same as in the previous section since they have the closed 
commutator algebra off-shell. Weyl weights and SO(8) representations 
of the fields are given in Table 2. 
Bosonic transformation laws other than the Weyl transformation 
are obvious from the index structure of the fields. 
Local supertransformations $\delta_Q$ and super Weyl transformations 
$\delta_S$ of the BLG fields are given by 
\begin{eqnarray}
\delta_Q X_i^I 
\A=\A \frac{1}{2\sqrt{2}} \bar{\epsilon} \Sigma^I \Psi_i, \nonu
\delta_Q \Psi_i 
\A=\A \frac{1}{2\sqrt{2}} \gamma^\mu \bar{\Sigma}^I \epsilon 
\hat{D}_\mu X_i^I 
+ \frac{1}{12\sqrt{2}} f_i{}^{jkl} X_j^J X_k^K X_l^L 
\bar{\Sigma}^{JKL} \epsilon 
%
- \frac{2}{3} \bar{\Sigma}^{JKL} \epsilon E^{IJKL} X_i^I, \nonu
\delta_Q \tilde{A}_\mu^j{}_i 
\A=\A \frac{1}{2\sqrt{2}} \bar{\epsilon} \gamma_\mu \Sigma^I \Psi_k
X_l^I f^{klj}{}_i 
- \frac{1}{8} \bar{\epsilon} \Sigma^{KL} \psi_\mu 
X_k^K X_l^L f^{klj}{}_i
\label{blgsuper}
\end{eqnarray}
and 
\begin{equation}
\delta_S X_i^I = 0, \quad
\delta_S \Psi_i = - \frac{1}{2\sqrt{2}} 
\bar{\Sigma}^I \eta X_i^I, \quad
\delta_S \tilde{A}_\mu^j{}_i = 0. 
\label{blgsuperweyl}
\end{equation}
Here, $\hat{D}_\mu$ is a supercovariant derivative 
\begin{equation}
\hat{D}_\mu X_i^I = D_\mu X_i^I 
- \frac{1}{2\sqrt{2}} \bar{\psi}_\mu \Sigma^I \Psi_i. 
\end{equation}
The covariant derivative $D_\mu$ contains the 3-algebra gauge fields 
in addition to the spin connection and the SO(8) gauge fields:
\begin{eqnarray}
D_\mu X_i^I \A=\A \partial_\mu X_i^I 
+ B_\mu^{IJ} X_i^J - X_j^I \tilde{A}_\mu{}^j{}_i, \nonu
D_\mu \Psi_i \A=\A \left( \partial_\mu 
+ \frac{1}{4} \hat{\omega}_{\mu ab} \gamma^{ab} 
+ \frac{1}{4} B_\mu^{IJ} \bar{\Sigma}^{IJ} \right) \Psi_i 
- \Psi_j \tilde{A}_\mu{}^j{}_i. 
\end{eqnarray}
\par

The Lagrangian of the BLG theory coupled to the conformal supergravity 
background can be written as 
\begin{eqnarray}
{\cal L} \A=\A {\cal L}_0 + {\cal L}_1 + {\cal L}_2, \nonu
{\cal L}_0 
\A=\A - \frac{1}{2} e D_\mu X^{iI} D^\mu X_i^I 
- \frac{1}{2} e \bar{\Psi}^i \gamma^\mu D_\mu \Psi_i 
+ \frac{1}{4} e f^{ijkl} \bar{\Psi}_i \bar{\Sigma}^{IJ} \Psi_j 
X_k^I X_l^J - eV \nonu
\A\A \mbox{} + \frac{1}{2} \epsilon^{\mu\nu\rho} \left( 
f^{ijkl} A_{\mu ij} \partial_\nu A_{\rho kl} 
+ \frac{2}{3} f^{kli}{}_p f^{mnpj} A_{\mu ij} A_{\nu kl} 
A_{\rho mn} \right), \nonu
{\cal L}_1 \A=\A \frac{1}{2\sqrt{2}} e \bar{\psi}_\mu \gamma^\nu
\gamma^\mu \Sigma^I \Psi^i D_\nu X_i^I 
+ \frac{1}{12\sqrt{2}} e \bar{\psi}_\mu \gamma^\mu 
\Sigma^{JKL} \Psi_i f^{ijkl} X_j^J X_k^K X_l^L \nonu
\A\A \mbox{} - \frac{1}{16} e \bar{\psi}_\mu \gamma^{\mu\nu\rho} 
\Sigma^{IJ} \psi_\nu X^{iI} D_\rho X_i^J 
- \frac{1}{4\sqrt{2}} e \bar{\Psi}^i \gamma^{\mu\nu} 
\bar{\Sigma}^I D_\mu \psi_\nu X_i^I 
- \frac{1}{16} e R X^2 \nonu
\A\A \mbox{} + \frac{1}{32} e \bar{\psi}_\mu \gamma^{\mu\nu\rho} 
D_\nu \psi_\rho X^2 
- \frac{1}{192} e \bar{\psi}_\mu \gamma^{\mu\nu} 
\Sigma^{IJKL} \psi_\nu f^{ijkl} X_i^I X_j^J X_k^K X_l^L, \nonu 
{\cal L}_2 \A=\A -4 e \bar{\lambda}^I \Psi^i X_i^I 
+ \frac{8}{3\sqrt{2}} e E^{IJKL} f^{ijkl} X_i^I X_j^J X_k^K X_l^L \nonu
\A\A \mbox{} - \frac{2}{3} e \bar{\psi}_\mu \gamma^\mu 
\Sigma^{JKL} \Psi^i E^{IJKL} X_i^I 
+ \sqrt{2} e D^{IJ} X^{iI} X_i^J \nonu
\A\A \mbox{} + \frac{1}{\sqrt{2}} e \bar{\psi}_\mu \gamma^\mu 
\Sigma^I \lambda^J X^{iI} X_i^J 
+ \frac{1}{6\sqrt{2}} e \bar{\psi}_\mu \gamma^{\mu\nu} \Sigma^{IKLM} 
\psi_\nu E^{JKLM} X^{iI} X_i^J \nonu
\A\A \mbox{} - \frac{4}{3} e E^{IJKL} E^{IJKL} X^2 
+ \frac{1}{6\sqrt{2}} e E^{IJKL} \bar{\Psi}^i \bar{\Sigma}^{IJKL} 
\Psi_i \nonu
\A\A \mbox{} + \frac{1}{32} e 
\bar{\psi}_\mu \gamma^\nu \gamma^\mu \psi_\nu \bar{\Psi}^i \Psi_i 
+ \frac{1}{128} e \bar{\psi}_\mu ( \gamma^{\mu\nu}{}_\rho
- g^{\mu\nu} \gamma_\rho ) \Sigma^{IJ} \psi_\nu 
\bar{\Psi}^i \gamma^\rho \bar{\Sigma}^{IJ} \Psi_i. 
\label{lagrangian}
\end{eqnarray}
Here, $R$ is the scalar curvature made from the spin connection 
$\hat{\omega}_{\mu ab}$ in (\ref{spinconnection}), and we have defined 
\begin{equation}
V = \frac{1}{12} f^{ijkl} f^{mnp}{}_l X_i^I X_j^J X_k^K 
X_m^I X_n^J X_p^K, \qquad
X^2 = X_i^I X^{iI}. 
\end{equation}
${\cal L}_0$ is an obvious generalization of the BLG Lagrangian 
to the one in a curved background with the spin connection 
(\ref{spinconnection}). 
${\cal L}_1$ consists of couplings to $e_\mu{}^a$, $\psi_\mu$ 
and $B_\mu^{IJ}$ of the conformal supergravity multiplet. 
These couplings appeared also in the Lagrangian 
in \cite{Gran:2008qx,Gran:2012mg}. 
${\cal L}_2$ contains couplings to other fields $\lambda^I$, $D^{IJ}$ 
and $E^{IJKL}$ of the conformal supergravity multiplet, 
and four-Fermi terms. 
The Lagrangian (\ref{lagrangian}) is invariant under all the local 
transformations up to a total divergence. 
\par

We found the fermionic transformations (\ref{blgsuper}), 
(\ref{blgsuperweyl}) and the Lagrangian (\ref{lagrangian})
by adding all possible terms which are invariant under the bosonic 
transformations with unknown coefficients and then fixing those 
coefficients such that the Lagrangian is invariant under the 
fermionic transformations. Finally, the complete invariance of the 
Lagrangian was shown. 
To show cancellations of terms cubic and higher in fermionic fields 
in $\delta_Q {\cal L}$ and $\delta_S {\cal L}$ we have used 
the identities  
\begin{eqnarray}
\A\A \gamma_a \Sigma^{IJKL} \psi_{[\rho} 
\bar{\psi}_\mu \gamma^a \psi_{\nu]} 
+ \gamma_a \psi_{[\rho} \bar{\psi}_\mu \gamma^a 
\Sigma^{IJKL} \psi_{\nu]} 
- 6 \Sigma^{[IJ} \psi_{[\rho} \bar{\psi}_\mu \Sigma^{KL]} \psi_{\nu]}
= 0, \nonu
\A\A \gamma^a \Sigma^{IJKL} \psi_{[\rho} 
\bar{\psi}_\mu \Sigma^{KL} \psi_{\nu]} 
+ \Sigma^{KL} \psi_{[\rho} \bar{\psi}_\mu \gamma^a 
\Sigma^{IJKL} \psi_{\nu]} \nonu
\A\A \mbox{} \qquad\qquad\qquad\qquad\qquad 
+ 6 \gamma^a \psi_{[\rho} \bar{\psi}_\mu \Sigma^{IJ} \psi_{\nu]} 
+ 6 \Sigma^{IJ} \psi_{[\rho} \bar{\psi}_\mu \gamma^a \psi_{\nu]} 
= 0, \nonu
\A\A \Sigma^{IJ} \psi_{[\rho} \bar{\psi}_\mu \Sigma^{IJ} \psi_{\nu]} 
+ 8 \gamma_a \psi_{[\rho} \bar{\psi}_\mu \gamma^a \psi_{\nu]} 
= 0, \nonu
\A\A \gamma_a \Sigma^{IJKL} \psi_{[\rho} \bar{\psi}_\mu \gamma^a 
\Sigma^{IJKL} \psi_{\nu]} 
+ 240 \gamma_a \psi_{[\rho} \bar{\psi}_\mu \gamma^a \psi_{\nu]} 
= 0, \nonu
\A\A \Sigma^{K(I} \psi_{[\rho} \bar{\psi}_\mu \Sigma^{J)K} \psi_{\nu]} 
= \delta^{IJ} \gamma_a \psi_{[\rho} \bar{\psi}_\mu \gamma^a \psi_{\nu]}, 
\end{eqnarray}
which can be proved by using the Fierz identities given in Appendix B. 
\par

The commutators of the local transformations now become 
\begin{eqnarray}
[ \delta_Q(\epsilon_1), \delta_Q(\epsilon_2) ] 
\A=\A \delta_G(\xi) + \delta_L(\lambda) + \delta_g(\zeta) 
+ \delta_{g3}(\tilde{\Lambda}) 
+ \delta_Q(\epsilon') + \delta_S(\eta'), \nonu
[ \delta_Q(\epsilon), \delta_S(\eta) ] 
\A=\A \delta_W(\Lambda) + \delta_L(\lambda') + \delta_g(\zeta') 
+ \delta_S(\eta''), \nonu
[ \delta_S(\eta_1), \delta_S(\eta_2) ] 
\A=\A 0.
\label{commutator3}
\end{eqnarray}
The transformation parameters on the right-hand sides are 
given by (\ref{rhsparameters}) and 
\begin{equation}
\tilde{\Lambda}^j{}_i 
= - \xi^\mu \tilde{A}_\mu{}^j{}_i 
- \frac{1}{8} \bar{\epsilon}_2 \Sigma^{KL} \epsilon_1 
X_k^K X_l^L f^{klj}{}_i
\end{equation}
for the 3-algebra gauge transformation.  
To show these commutation relations one has to use the 
field equations of the BLG fields $\Psi_i$ and $\tilde{A}_\mu{}^j{}_i$ 
derived from the Lagrangian (\ref{lagrangian}). 
Thus the algebra closes only on-shell on the BLG fields 
as in the original BLG theory in a flat background. 


%
\newsection{Supercurrent multiplet of the BLG theory}
As an application of the results in the previous sections let us 
obtain a supercurrent multiplet of the BLG theory in a flat background. 
The supercurrent multiplet is a supermultiplet which contains the 
energy-momentum tensor and the supercurrent in addition to other 
quantities \cite{Ferrara:1974pz,Bergshoeff:1980is}. 
\par

We consider the BLG theory in a flat conformal supergravity background 
\begin{equation}
e_\mu{}^a = \delta_\mu^a, \quad
\psi_\mu = B_\mu^{IJ} = \lambda^I = D^{IJ} = E^{IJKL} = 0. 
\label{flatbackground}
\end{equation}
This background is preserved by a part of the local transformations 
discussed in section 3. Such transformations form a supergroup 
OSp($8|4$), which contains SO($2,3$) $\times$ SO(8) as a bosonic
subgroup. The Lagrangian (\ref{lagrangian}) in this background 
is invariant under OSp($8|4$) transformations \cite{Bandres:2008vf}. 
OSp($8|4$) consists of conformal transformation $\delta_C$, 
superconformal transformation $\delta_{SC}$ and global SO(8) 
transformation $\delta_{\rm SO(8)}$ given by 
\begin{eqnarray}
\delta_C(\xi) \A=\A \delta_G(\xi) + \delta_L(\lambda) 
+ \delta_W(\Lambda), \qquad
\lambda_{\mu\nu} = - \partial_{[\mu} \xi_{\nu]}, \ \ 
\Lambda = - \frac{1}{3} \partial_\mu \xi^\mu, \nonu
\delta_{SC}(\epsilon) \A=\A \delta_Q(\epsilon) + \delta_S(\eta), \qquad
\eta = - \frac{1}{3} \gamma^\mu \partial_\mu \epsilon, \nonu
\delta_{\rm SO(8)}(\zeta) \A=\A \delta_g(\zeta), 
\label{osp}
\end{eqnarray}
where the transformation parameters $\xi^\mu$, $\epsilon$ and 
$\zeta^{IJ}$ satisfy 
\begin{equation}
\partial_\mu \xi_\nu + \partial_\nu \xi_\mu 
= \frac{2}{3} \, \eta_{\mu\nu} \partial_\rho \xi^\rho, \quad
\left( \partial_\mu - \frac{1}{3} \gamma_\mu \gamma^\nu 
\partial_\nu \right) \epsilon = 0, \quad
\partial_\mu \zeta^{IJ} = 0.
\label{ckeq}
\end{equation}
Solutions of the first two equations are conformal Killing vectors 
and conformal Killing spinors, respectively. 
The general form of conformal Killing spinors is 
\begin{equation}
\epsilon = \alpha + x^\mu \gamma_\mu \beta,
\end{equation}
where $\alpha$ and $\beta$ are arbitrary constant spinors. 
\par

The supercurrent multiplet of the BLG theory can be obtained by 
computing derivatives of the Lagrangian (\ref{lagrangian}) 
with respect to the conformal supergravity fields and taking 
the flat background (\ref{flatbackground}). 
We find the supercurrent multiplet as 
\begin{eqnarray}
T_{\mu\nu} \A=\A D_\mu X^{iI} D_\nu X_i^I 
- \frac{1}{2} \eta_{\mu\nu} D_\rho X^{iI} D^\rho X_i^I 
- \frac{1}{8} ( \partial_\mu \partial_\nu - \eta_{\mu\nu} \partial^2 ) 
( X^{iI} X_i^I ) \nonu
\A\A \mbox{} + \frac{1}{2} \bar{\Psi}^i \gamma_{(\mu} D_{\nu)} \Psi_i 
- \frac{1}{12} \eta_{\mu\nu} f^{ijkl} f^{mnp}{}_l X_i^I X_j^J X_k^K
X_m^I X_n^J X_p^K, \nonu
S^\mu \A=\A \frac{1}{2\sqrt{2}} \gamma^\nu \gamma^\mu \Sigma^I
\Psi^i D_\nu X_i^I 
+ \frac{1}{4\sqrt{2}} \partial_\nu ( \gamma^{\mu\nu} \Sigma^I 
\Psi^i X_i^I ) \nonu
\A\A \mbox{} + \frac{1}{12\sqrt{2}} f^{ijkl} \gamma^\mu \Sigma^{JKL} \Psi_i
X_j^J X_k^K X_l^L, \nonu
J_\mu^{IJ} \A=\A X^{i[I} D_\mu X_i^{J]} 
- \frac{1}{8} \bar{\Psi}^i \gamma_\mu \bar{\Sigma}^{IJ} \Psi_i, \nonu
R^I \A=\A \Psi^i X_i^I 
- \frac{1}{8} \bar{\Sigma}^I \Sigma^J \Psi^i X_i^J, \nonu
M^{IJ} \A=\A X_i^I X^{iJ} - \frac{1}{8} \delta^{IJ} X_i^K X^{iK}, \nonu
N^{IJKL} \A=\A \frac{1}{2} f^{ijkl} \left( X_i^I X_j^J X_k^K X_l^L 
+ \frac{1}{4!} \epsilon^{IJKLMNPQ} X_i^M X_j^N X_k^P X_l^Q \right) 
+ \frac{1}{16} \bar{\Psi}^i \bar{\Sigma}^{IJKL} \Psi_i, \nonu
\label{supercurrent}
\end{eqnarray}
where the covariant derivative $D_\mu$ contains only the 3-algebra 
gauge field $\tilde{A}_\mu{}^j{}_i$. 
$T_{\mu\nu}$, $S^\mu$ and $J_\mu^{IJ}$ are the energy-momentum tensor, 
the supercurrent and the SO(8) current, respectively. 
They satisfy conservation laws and ($\gamma$-)traceless conditions 
\begin{eqnarray}
\partial_\mu T^{\mu\nu} \A=\A 0, \qquad
\partial_\mu S^\mu = 0, \qquad
\partial_\mu J^{\mu IJ} = 0, \nonu
T_\mu{}^\mu \A=\A 0, \qquad
\gamma_\mu S^\mu = 0. 
\label{conservation}
\end{eqnarray}
From these quantities we can construct the OSp($8|4$) charges 
\begin{equation}
Q_T[\xi] = \int d^3 x \, \xi^\nu T_\nu{}^0, \quad
Q_S[\epsilon] = \int d^3 x \, \bar{\epsilon} S^{0}, \quad
Q_J[\zeta] = \int d^3 x \, \zeta^{IJ} J^{0IJ}, 
\label{conservedcharge}
\end{equation}
where $\xi^\mu$, $\epsilon$ and $\zeta^{IJ}$ are solutions of 
(\ref{ckeq}). By (\ref{conservation}) and (\ref{ckeq}) these 
charges are conserved. 
$R^I$, $M^{IJ}$, $N^{IJKL}$ are quantities corresponding to 
the fields $\lambda^I$, $D^{IJ}$, $E^{IJKL}$, respectively. 
$R^I$, $M^{IJ}$ satisfy ($\Sigma$-)traceless conditions 
$\Sigma^I R^I = 0$, $M^{II} = 0$ and $N^{IJKL}$ is self-dual. 
\par

Under the OSp($8|4$) transformations in (\ref{osp}) 
these quantities transform into themselves. In particular, 
their superconformal transformations are 
\begin{eqnarray}
\delta_{SC} T_{\mu\nu} \A=\A \frac{1}{2} \bar{\epsilon} 
\gamma_{\rho(\mu} \partial^\rho S_{\nu)}
+ 2 \partial_{(\mu} \bar{\epsilon} S_{\nu)}, \nonu
\delta_{SC} S_\mu \A=\A \frac{1}{4} \gamma^\nu \epsilon T_{\mu\nu} 
- \frac{1}{16} \gamma^\rho \gamma_\mu{}^\sigma \Sigma^{IJ} \epsilon
\partial_\sigma J_\rho^{IJ}
- \frac{1}{16} \Sigma^{IJ} ( \gamma_\mu \partial_\nu 
+ 3 \gamma_\nu \partial_\mu ) \epsilon J^{\nu IJ}, \nonu
\delta_{SC} J_\mu^{IJ} \A=\A \frac{1}{4} \bar{\epsilon} \Sigma^{IJ} S_\mu 
+ \frac{1}{2\sqrt{2}} \partial_\nu ( \bar{\epsilon} 
\gamma_\mu{}^\nu \Sigma^{[I} R^{J]} ), \nonu
\delta_{SC} R^I \A=\A \frac{1}{2\sqrt{2}} \gamma^\mu \bar{\Sigma}^J
\epsilon J_\mu^{IJ} 
- \frac{1}{16\sqrt{2}} \gamma^\mu \bar{\Sigma}^I \Sigma^{KL} \epsilon
J_\mu^{KL} 
+ \frac{1}{12\sqrt{2}} \bar{\Sigma}^{JKL} \epsilon N^{IJKL} \nonu
\A\A \mbox{} + \frac{1}{4\sqrt{2}} \gamma^\mu \bar{\Sigma}^J \epsilon
\partial_\mu M^{IJ} 
+ \frac{1}{6\sqrt{2}} \gamma^\mu \bar{\Sigma}^J \partial_\mu 
\epsilon M^{IJ}, \nonu
\delta_{SC} M^{IJ} \A=\A \frac{1}{\sqrt{2}} \bar{\epsilon} 
\Sigma^{(I} R^{J)}, \nonu
\delta_{SC} N^{IJKL} \A=\A - \frac{1}{16\sqrt{2}} \bar{\epsilon} 
\gamma^\mu \Sigma^M \bar{\Sigma}^{IJKL} \partial_\mu R^M 
- \frac{1}{48\sqrt{2}} \partial_\mu \bar{\epsilon} \gamma^\mu 
\Sigma^M \bar{\Sigma}^{IJKL} R^M.
\end{eqnarray}
We see that $T_{\mu\nu}$, $S_\mu$ and $J_\mu^{IJ}$ transform 
into themselves except that $R^I$ appears in $\delta_{SC} J_\mu^{IJ}$.  
However, this $R^I$ term does not contribute to transformations of 
the conserved charges (\ref{conservedcharge}). 
Indeed, we find that they transform as 
\begin{eqnarray}
\delta_{SC}(\epsilon_1) Q_T[\xi_2] \A=\A Q_S[\epsilon_3], \nonu 
\delta_{SC}(\epsilon_1) Q_S[\epsilon_2] 
\A=\A Q_T[\xi_3] + Q_J[\zeta_3], \nonu 
\delta_{SC}(\epsilon_1) Q_J[\zeta_2] \A=\A Q_S[\epsilon'_3], 
\end{eqnarray}
where the transformation parameters on the right-hand sides are 
\begin{eqnarray}
\epsilon_3 \A=\A \xi_2^\mu \partial_\mu \epsilon_1 
+ \frac{1}{4} \partial_\mu \xi_{2\nu} \gamma^{\mu\nu} \epsilon_1 
- \frac{1}{6} \partial_\mu \xi_2^\mu \epsilon_1, \qquad
\epsilon'_3 = - \frac{1}{4} \zeta_2^{IJ} \Sigma^{IJ} \epsilon_1, \nonu
\xi^\mu_3 \A=\A \frac{1}{4} \bar{\epsilon}_2 \gamma^\mu \epsilon_1, \qquad
\zeta^{IJ}_3 = - \frac{1}{24} ( \bar{\epsilon}_2 \Sigma^{IJ} 
\gamma^\mu \partial_\mu \epsilon_1 
- \bar{\epsilon}_1 \Sigma^{IJ} \gamma^\mu \partial_\mu \epsilon_2 ). 
\end{eqnarray}
It can be shown that these parameters satisfy (\ref{ckeq}), i.e., 
$\epsilon_3$, $\epsilon'_3$ are conformal Killing spinors, 
$\xi_3^\mu$ is a conformal Killing vector and 
$\zeta_3^{IJ}$ is a constant. 
The conserved charges (\ref{conservedcharge}) are transformed into 
themselves also by the conformal and SO(8) transformations.

%
\newsection{Discussion}
In this paper we obtained local transformation laws 
of the $D=3$, ${\cal N}=8$ conformal supergravity. 
There are two cases 
depending on the parameter $\eta=\pm 1$ appearing 
in the (anti) self-duality conditions on the scalar fields. 
In both cases the commutator algebra of the local transformations 
closes off-shell on the conformal supergravity fields. 
In the $\eta=+1$ case we coupled the conformal supergravity 
to the BLG theory and obtained local transformation laws and 
an invariant Lagrangian of the coupled theory. 
The commutator algebra closes also on the BLG fields when their 
field equations are used. 
\par

On the other hand, in the $\eta=-1$ case we did not succeed in 
coupling the conformal supergravity to the BLG theory. 
We found a difficulty in a commutator of local supertransformations. 
Assuming that the local supertransformation laws of the BLG fields 
are given by 
\begin{equation}
\delta_Q \Psi_i 
= \frac{1}{2\sqrt{2}} \gamma^\mu \bar{\Sigma}^I \epsilon 
\hat{D}_\mu X_i^I 
+ \frac{1}{12\sqrt{2}} f_i{}^{jkl} X_j^J X_k^K X_l^L 
\bar{\Sigma}^{JKL} \epsilon 
+ \frac{1}{2} \bar{\Sigma}^I \epsilon E^{IJ} X_i^J
\label{superpsi}
\end{equation}
and $\delta_Q X_i^I$, $\delta_Q \tilde{A}_\mu{}^j{}_i$ in 
(\ref{blgsuper}), we found an extra term in the commutation relation 
\begin{equation}
[ \delta_Q(\epsilon_1), \delta_Q(\epsilon_2) ] X_i^I
= \cdots - \frac{1}{2\sqrt{2}} \bar{\epsilon}_2 \Sigma^{K(I} 
\epsilon_1 E^{J)K} X_i^J, 
\end{equation}
where $\cdots$ denote the expected local transformation terms 
in (\ref{commutator3}). 
Therefore, this commutator does not close. 
A change of the coefficient in the last term of (\ref{superpsi})
or an addition of terms of the form 
$\bar{\Sigma}^I \epsilon X^I X^J X^J$ 
to (\ref{superpsi}) does not improve the situation. 
Another puzzling point is on the supercurrent multiplet of 
the BLG theory in a flat background. 
If one could find a coupling of the $\eta=-1$ conformal 
supergravity, one would obtain a supercurrent multiplet different 
from (\ref{supercurrent}). In particular, $M^{IJ}$, $N^{IJKL}$ would be 
replaced by something like $M^{IJKL}$, $N^{IJ}$. 
Since these multiplets are those of the original BLG theory 
without coupling to the conformal supergravity, 
we would obtain two different supercurrent multiplets 
for the same one theory. 
These observations seem to suggest that it is not possible to 
couple the $\eta=-1$ off-shell conformal supergravity to the BLG theory.  
\par

The $D=3$, ${\cal N}=8$ conformal supergravity and its coupling to 
the BLG theory were previously discussed 
in \cite{Gran:2008qx,Gran:2012mg,Howe:1995zm,Cederwall:2011pu} in both 
of the component field formulation and the superspace formulation. 
Let us compare these works with ours. 
In the component field formulation \cite{Gran:2008qx,Gran:2012mg} 
the on-shell conformal supergravity 
multiplet, which consists of a dreibein $e_\mu{}^a$, Rarita-Schwinger 
fields $\psi_\mu^\alpha$ and SO(8) gauge fields $B_\mu^{IJ}$, was used. 
The Lagrangian and the local supertransformations of these fields 
were obtained. The conformal supergravity multiplet has a kinetic 
term in the Lagrangian and is dynamical. 
The commutator algebra closes only on-shell, i.e., when field 
equations of the conformal supergravity fields are used. 
In the superspace 
formulation \cite{Howe:1995zm,Cederwall:2011pu,Gran:2012mg} 
the conformal supergravity multiplet is represented by superfields. 
The superfield of central importance is the ``super Cotton tensor'', 
which contains the Cotton tensor for the gravitational field, 
its superpartner (Cottino) for the Rarita-Schwinger fields 
and the field strength of the SO(8) gauge fields as component fields. 
In addition, the super Cotton tensor also contains spinor and 
scalar fields $\lambda^{\alpha\beta\gamma}$,
$D^{\alpha\beta\gamma\delta}$, 
$E^{\alpha\beta\gamma\delta}$. Its leading component field is 
the scalar field $E^{\alpha\beta\gamma\delta}$, which is either 
self-dual or anti self-dual. 
Therefore, this formulation corresponds to 
the off-shell conformal supergravity multiplet as in our approach. 
However, when it is coupled to the BLG matter theory, 
the field equations of the matter fields were given 
only when the conformal supergravity fields are on-shell, i.e., they 
satisfy their field equations \cite{Gran:2012mg}. 
In particular, the spinor and scalar fields are not independent 
fields but are expressed in terms of the matter fields as 
determined by (4.25), (4.26) of \cite{Gran:2012mg}. 
Then, the superspace formulation reduces to the on-shell 
component field formulation. 
\par

To compare these results with ours we first 
suppose that our results on the off-shell conformal 
supergravity multiplet in section 3 correspond to 
a component field form of the superspace formulation 
in \cite{Howe:1995zm,Cederwall:2011pu,Gran:2012mg} 
although we have not shown it explicitly. 
As for the coupling to the BLG theory we note that in the 
superspace formulation \cite{Gran:2012mg} the coupling was 
studied only when $E^{\alpha\beta\gamma\delta}$ is self-dual and 
not anti self-dual. This corresponds to our $\eta=-1$ case 
as can be seen by comparing (D.13), (D.14) of \cite{Gran:2012mg} 
and our (\ref{newde}) with $D$ and $E$ interchanged for $\eta=-1$. 
Although we could not find a coupled theory for $\eta=-1$, 
the supertransformations of the conformal supergravity multiplet 
were obtained and are given in (\ref{super2}). 
Comparing the supertransformations of the fields $e_\mu{}^a$, 
$\psi_\mu$, $B_\mu^{IJ}$ in (\ref{super2}) and those in 
(2.20)--(2.22) of \cite{Gran:2012mg} we see that they are the same 
if we assume\footnote{The right-hand sides of these equations are 
proportional to $R^I$ and $M^{IJ}$ in the supercurrent multiplet 
of the $\eta=+1$ theory (\ref{supercurrent}), respectively. 
It seems natural then to also assume that $D^{IJKL}$ is proportional 
to $N^{IJKL}$ in (\ref{supercurrent}).} 
\begin{eqnarray}
\lambda^I \A=\A \frac{1}{8} \, g \left( \Psi^i X_i^I 
- \frac{1}{8} \bar{\Sigma}^I \Sigma^J \Psi^i X_i^J \right), \nonu
E^{IJ} \A=\A \frac{1}{4\sqrt{2}} \, g \left( 
X_i^I X^{iJ} - \frac{1}{8} \delta^{IJ} X_i^K X^{iK} \right),  
\label{ansatz}
\end{eqnarray}
where $g$ is a conformal gravitational coupling constant used in 
\cite{Gran:2012mg}. Furthermore, (\ref{ansatz}) appear as field 
equations for the conformal supergravity multiplet in the 
superspace formulation ((4.25), (4.26) of \cite{Gran:2012mg}). 
Therefore, our results for $\eta=-1$ are consistent with those 
of \cite{Gran:2008qx,Gran:2012mg} as far as the transformations 
of the conformal supergravity multiplet are concerned. 
Although we could not find a coupled theory for the off-shell 
conformal supergravity multiplet, it is possible to construct 
a coupled theory for the on-shell multiplet as was done 
in \cite{Gran:2008qx,Gran:2012mg}. 
\par

A coupling of the BLG theory to the $\eta=+1$ conformal 
supergravity multiplet was not considered in the superspace 
formulation in \cite{Gran:2012mg}. 
It would be interesting to study a coupling to the $\eta=+1$ 
multiplet in the superspace formulation 
and compare it with our results in section 4. 
In connection with this it would be also interesting to see 
whether one can construct a theory in which 
the $\eta=+1$ conformal supergravity multiplet is dynamical 
and can eliminate the auxiliary spinor and scalar fields 
in terms of the matter fields by using field equations 
to obtain an on-shell multiplet. If such a theory is possible, 
the second equation of (\ref{ansatz}) will be replaced 
by an equation for $E^{IJKL}$. One may guess that it is  
something like $E^{IJKL} \sim g N^{IJKL} (X^2)^{-1}$, 
where $N^{IJKL}$ is given in (\ref{supercurrent}) and 
the factor $(X^2)^{-1}$ is needed to match the Weyl weight. 
Then, on-shell supertransformations will have different forms 
from those of \cite{Gran:2008qx,Gran:2012mg}. 
\par

Finally, $D=3$ conformal supergravity multiplets with 
lower ${\cal N}$ supersymmetries can be obtained from 
the ${\cal N}=8$ multiplet given in section 3 
by consistent truncations, i.e., by setting some of the fields 
to zero in such a way that a part of supersymmetries is preserved. 
Their local transformation laws are easily obtained from 
those of the ${\cal N}=8$ multiplet. For instance, 
a truncation to ${\cal N}=6$ can be obtained as follows. 
In section 3 we used SO(8) negative chirality indices 
$\alpha, \beta, \gamma, \cdots = 1,2,\cdots, 8$ for 
the ${\cal N}=8$ conformal supergravity multiplet as given 
in Table 1 and the parameters of the super and super Weyl 
transformations $\epsilon^\alpha$, $\eta^\alpha$. 
Alternatively, 
we can consistently replace all these indices by SO(8) vector 
indices $I, J, K, \cdots =1,2,\cdots,8$. 
Then, the conformal supergravity multiplet and the transformation 
parameters become  
$e_\mu{}^a$, $\psi_\mu^I$, $B_\mu^{IJ}$, $\lambda^{IJK}$, 
$D^{IJKL}$, $E^{IJKL}$ and $\epsilon^I$, $\eta^I$. 
The truncation to ${\cal N}=6$ is given by setting 
$\psi_\mu^{I'}=0$, $B_\mu^{IJ'}=0$, 
$\lambda^{IJK'}=0$, $D^{IJKL'}=0$, 
$E^{IJKL'}=0$, where 
$I, J, K, \cdots = 1,2,\cdots,6$ and 
$I', J', K', \cdots = 7,8$. 
These conditions are invariant under the super and super Weyl
transformations (\ref{supertransformation}), (\ref{superweyl}) 
when $\epsilon^{I'}=0$, $\eta^{I'}=0$. 
The remaining independent fields are 
\begin{equation}
e_\mu{}^a, \quad \psi_\mu^I, \quad B_\mu^{IJ}, 
\quad B_\mu^{78}, \quad \lambda^{IJK}, 
\quad \lambda^{I78}, \quad 
D^{IJKL}, \quad
E^{IJKL}, 
\label{n6}
\end{equation}
where $I, J, K, \cdots = 1,2,\cdots,6$. 
The fields $D^{IJ78}$, $E^{IJ78}$ are related to 
$D^{IJKL}$, $E^{IJKL}$ in (\ref{n6}) 
by the self-duality conditions (\ref{scalarduality2}).
The SO(8) gauge transformation is reduced to SO(6) $\times$ U(1), 
whose gauge fields are $B_\mu^{IJ}$ and $B_\mu^{78}$ in (\ref{n6}). 
The fields in (\ref{n6}) constitute the ${\cal N}=6$ conformal 
supergravity multiplet as given by the superspace formulation 
\cite{Howe:1995zm,Gran:2012mg}. 
The ${\cal N}=6$ multiplet is interesting since it can be coupled 
to the ABJM theory \cite{Aharony:2008ug} of multiple M2-branes. 
A coupling of the dynamical ${\cal N}=6$ conformal supergravity 
to the ABJM theory was already studied in the on-shell component 
field formulation \cite{Chu:2009gi,Chu:2010fk} and in the off-shell 
superspace formulation \cite{Gran:2012mg}. 
For ${\cal N} \leq 4$, off-shell conformal supergravities and 
their couplings to matter multiplets were studied in the superspace 
formulation in \cite{Kuzenko:2011xg}. 
We can use the lower ${\cal N}$ off-shell conformal supergravity 
multiplets obtained by the truncations from ${\cal N}=8$ to study 
their coupling to matter multiplets in terms of component fields. 

\bigskip

\noindent {\bf{\large Acknowledgments}}

M.N. would like to thank Theoretical High Energy Physics Group 
of Bielefeld University, where part of this work was done, 
for hospitality.

\begin{appendix}

%
\def\numberbysectiona{\@addtoreset{equation}{section}
\def\theequation{A.\arabic{equation}}}
\numberbysectiona
\setcounter{section}{0}
\setcounter{equation}{0}
\setcounter{subsection}{0}
\setcounter{footnote}{0}

\newsection{Notations and Conventions}
Three-dimensional world and local Lorentz indices 
are denoted by $\mu, \nu, \cdots = 0,1,2$ and 
$a, b, \cdots = 0,1,2$, respectively. 
A flat metric is $\eta_{ab}={\rm diag}(-1,+1,+1)$ and 
the antisymmetric symbol $\epsilon^{abc}$ is chosen as 
$\epsilon^{012}=+1$. 
Symmetrization and antisymmetrization of indices with weight one 
are denoted as $(ab\cdots)$ and $[ab\cdots]$, respectively. 
Three-dimensional gamma matrices are denoted by $\gamma^a$. 
Their antisymmetrized products $\gamma^{ab}=\gamma^{[a}\gamma^{b]}$, 
$\gamma^{abc}=\gamma^{[a} \gamma^b \gamma^{c]}$ satisfy 
\begin{equation}
\gamma^{abc} = - \epsilon^{abc}, \qquad
\gamma^{ab} = - \epsilon^{abc} \gamma_c, \qquad
\gamma^a = \frac{1}{2} \epsilon^{abc} \gamma_{bc}. 
\end{equation}
\par

There are three eight-dimensional representations of SO(8): 
a vector representation ${\bf 8}_v$, a negative chirality Weyl 
spinor representation ${\bf 8}_s$ and a positive chirality Weyl 
spinor representation ${\bf 8}_c$. 
Indices of these representations are denoted by 
$I,J,\cdots=1,2,\cdots,8$; 
$\alpha, \beta, \cdots =1,2,\cdots,8$ and 
$\dot{\alpha}, \dot{\beta}, \cdots =1,2,\cdots,8$, respectively. 
For these indices upper and lower indices are not distinguished. 


%
\def\numberbysectionb{\@addtoreset{equation}{section}
\def\theequation{B.\arabic{equation}}}
\numberbysectionb

\newsection{SO(8) identities}
In this Appendix we summarize some properties of the matrices 
$\Sigma^I$ and $\bar{\Sigma}^I =(\Sigma^I)^T$ introduced in 
(\ref{so8gamma}). From the anti-commutation relation of the SO(8) 
gamma matrices $\{ \Gamma^I, \Gamma^J \} = 2 \delta^{IJ}$ they satisfy 
\begin{eqnarray}
(\Sigma^I)_{\alpha\dot{\gamma}} (\Sigma^J)_{\beta\dot{\gamma}} 
+ (\Sigma^J)_{\alpha\dot{\gamma}} (\Sigma^I)_{\beta\dot{\gamma}} 
\A=\A 2 \delta^{IJ} \delta_{\alpha\beta}, \nonu
(\Sigma^I)_{\gamma\dot{\alpha}} (\Sigma^J)_{\gamma\dot{\beta}} 
+ (\Sigma^J)_{\gamma\dot{\alpha}} (\Sigma^I)_{\gamma\dot{\beta}} 
\A=\A 2 \delta^{IJ} \delta_{\dot{\alpha}\dot{\beta}}. 
\end{eqnarray}
They also satisfy 
\begin{equation}
(\Sigma^I)_{\alpha\dot{\alpha}} (\Sigma^I)_{\beta\dot{\beta}} 
+ (\Sigma^I)_{\alpha\dot{\beta}} (\Sigma^I)_{\beta\dot{\alpha}} 
= 2 \delta_{\alpha\beta} \delta_{\dot{\alpha}\dot{\beta}}
\end{equation}
as can be shown by using the Fierz identity. 
Similarity of these three equations is a consequence of the 
triality of three eight-dimensional representations of SO(8). 
We choose sign conventions of these matrices such that 
\begin{eqnarray}
(\Sigma^{[I_1})_{\alpha_1\dot{\alpha}_2} 
(\bar{\Sigma}^{I_2})_{\dot{\alpha}_2\alpha_3} 
(\Sigma^{I_3})_{\alpha_3\dot{\alpha}_4} \cdots 
(\bar{\Sigma}^{I_8]})_{\dot{\alpha}_8\alpha_9} 
\A=\A - \delta_{\alpha_1\alpha_9} \epsilon^{I_1\cdots I_8}, \nonu
(\Sigma^{I_1})_{[\alpha_1}{}^{\dot{\alpha}_2}
(\Sigma^{I_2} )_{\alpha_2}{}^{\dot{\alpha}_2}
(\Sigma^{I_2})_{\alpha_3}{}^{\dot{\alpha}_3} \cdots 
(\Sigma^{I_5})_{\alpha_8]}{}^{\dot{\alpha}_5}
\A=\A \delta^{I_1I_5} \epsilon_{\alpha_1\cdots \alpha_8}
\label{sigmaconvention}
\end{eqnarray}
are satisfied. 
\par

We denote antisymmetrized products of these matrices as 
\begin{eqnarray}
\Sigma^{IJ} \A=\A \Sigma^{[I} \bar{\Sigma}^{J]}, \quad
\Sigma^{IJK} = \Sigma^{[I} \bar{\Sigma}^{J} \Sigma^{K]}, \quad
\cdots, \nonu
\bar{\Sigma}^{IJ} \A=\A \bar{\Sigma}^{[I} \Sigma^{J]}, \quad
\bar{\Sigma}^{IJK} = \bar{\Sigma}^{[I} \Sigma^{J} \bar{\Sigma}^{K]},
\quad \cdots.
\end{eqnarray}
In general the leftmost matrix in $\Sigma^{IJ\cdots}$ is 
$\Sigma^I$ while that in $\bar{\Sigma}^{IJ\cdots}$ is $\bar{\Sigma}^I$. 
By the first equation of (\ref{sigmaconvention}) these matrices satisfy 
\begin{eqnarray}
\Sigma^{I_1\cdots I_4} \A=\A - \frac{1}{4!} \epsilon^{I_1\cdots I_8} 
\Sigma^{I_5\cdots I_8}, \nonu
\Sigma^{I_1\cdots I_5} 
\A=\A \frac{1}{3!} \epsilon^{I_1\cdots I_8} 
\Sigma^{I_6I_7I_8}, \nonu
\Sigma^{I_1\cdots I_6} 
\A=\A \frac{1}{2} \epsilon^{I_1\cdots I_8} 
\Sigma^{I_7I_8}, \nonu
\Sigma^{I_1\cdots I_7} 
\A=\A - \epsilon^{I_1\cdots I_8} \Sigma^{I_8}, \nonu
\Sigma^{I_1\cdots I_8} \A=\A - \epsilon^{I_1\cdots I_8} 
\end{eqnarray}
and similar identities for $\bar{\Sigma}^{IJ\cdots}$ 
with opposite signs on the right-hand sides. 
In particular, $\Sigma^{I_1\cdots I_4}$ and 
$\bar{\Sigma}^{I_1\cdots I_4}$ are anti self-dual and self-dual, 
respectively. Other useful identities are 
\begin{align}
\Sigma^M \bar{\Sigma}^M &= 8, &
\Sigma^{MN} \Sigma^{MN} &= -56, \nonu
\Sigma^M \bar{\Sigma}^{I} \Sigma^M &= -6 \Sigma^I, &
\Sigma^{MN} \Sigma^{I} \bar{\Sigma}^{MN} &= -28 \Sigma^{I}, \nonu
\Sigma^M \bar{\Sigma}^{IJ} \bar{\Sigma}^M &= 4 \Sigma^{IJ}, &
\Sigma^{MN} \Sigma^{IJ} \Sigma^{MN} &= -8 \Sigma^{IJ}, \nonu
\Sigma^M \bar{\Sigma}^{IJK} \Sigma^M &= -2 \Sigma^{IJK}, &
\Sigma^{MN} \Sigma^{IJK} \bar{\Sigma}^{MN} &= 4 \Sigma^{IJK}, \nonu
\Sigma^M \bar{\Sigma}^{IJKL} \bar{\Sigma}^M &= 0, &
\Sigma^{MN} \Sigma^{IJKL} \Sigma^{MN} &= 8 \Sigma^{IJKL}.
\end{align}
\par

Bilinears of two SO($1,2$) Majorana spinors $\psi_1^\alpha$ and 
$\psi_2^\alpha$ in the representation ${\bf 8}_s$ of SO(8) have 
symmetries 
\begin{align}
\bar{\psi}_1 \psi_2 &= \bar{\psi}_2 \psi_1, &
\bar{\psi}_1 \gamma^a \psi_2 &= - \bar{\psi}_2 \gamma^a \psi_1, \nonu
\bar{\psi}_1 \Sigma^{IJ} \psi_2 
&= - \bar{\psi}_2 \Sigma^{IJ} \psi_1, &
\bar{\psi}_1 \gamma^a \Sigma^{IJ} \psi_2 
&= \bar{\psi}_2 \gamma^a \Sigma^{IJ} \psi_1, \nonu
\bar{\psi}_1 \Sigma^{IJKL} \psi_2 
&= \bar{\psi}_2 \Sigma^{IJKL} \psi_1, &
\bar{\psi}_1 \gamma^a \Sigma^{IJKL} \psi_2 
&= - \bar{\psi}_2 \gamma^a \Sigma^{IJKL} \psi_1. 
\end{align}
The Fierz identity for spinors $\psi_1^\alpha,\ \cdots,\ \psi_4^\alpha$ 
can be written as 
\begin{eqnarray}
\bar{\psi}_1 \psi_2 \bar{\psi}_3 \psi_4 
\A=\A - \frac{1}{16} \left( \bar{\psi}_1 \psi_4 \bar{\psi}_3 \psi_2 
+ \bar{\psi}_1 \gamma^a \psi_4 
\bar{\psi}_3 \gamma_a \psi_2 \right) \nonu
\A\A \mbox{} + \frac{1}{32} \left( \bar{\psi}_1 \Sigma^{IJ} \psi_4 
\bar{\psi}_3 \Sigma^{IJ} \psi_2 
+ \bar{\psi}_1 \gamma^a \Sigma^{IJ} \psi_4 
\bar{\psi}_3 \gamma_a \Sigma^{IJ} \psi_2 \right) \nonu
\A\A \mbox{}
- \frac{1}{32 \cdot 4!} \left( \bar{\psi}_1 \Sigma^{IJKL} \psi_4 
\bar{\psi}_3 \Sigma^{IJKL}\psi_2 
+ \bar{\psi}_1 \gamma^a \Sigma^{IJKL} 
\psi_4 \bar{\psi}_3 \gamma_a \Sigma^{IJKL} \psi_2 \right). \nonu
\end{eqnarray}
\par

When totally antisymmetric tensors $S^{IJKL}$, $S'^{IJKL}$ are 
self-dual and $A^{IJKL}$, $A'^{IJKL}$ are anti self-dual, 
the following identities are satisfied: 
\begin{eqnarray}
S^{IJKL} A^{IJKL} \A=\A 0, \nonu
S^{IKLM} A^{JKLM} \A=\A S^{JKLM} A^{IKLM}, \nonu
S^{IJMN} A^{KLMN} + S^{KLMN} A^{IJMN} 
\A=\A - \frac{4}{3} S^{MNP[I} \delta^{J][L} A^{K]MNP}, \nonu
S^{KLM(I} S'^{J)KLM} 
\A=\A - \frac{1}{8} \delta^{IJ} S^{KLMN} S'^{KLMN}, \nonu
A^{KLM(I} A'^{J)KLM} 
\A=\A - \frac{1}{8} \delta^{IJ} A^{KLMN} A'^{KLMN}.
\end{eqnarray}

\end{appendix}

\end{document}